\documentclass[%
 reprint,longbibliography,preprintnumbers,
nofootinbib,
 amsmath,amssymb,
 aps,
pre,
]{revtex4-2}
\usepackage{graphicx}
\usepackage[utf8]{inputenc}
\usepackage{flushend}
\usepackage{dcolumn}
\usepackage{bm}
\usepackage{balance}

\usepackage{xcolor} 
\usepackage[normalem]{ulem} %

\usepackage[normalem]{ulem}

\usepackage{verbatim}
\usepackage{color,ulem}
\usepackage[english]{babel}

\usepackage[utf8]{inputenc}

\input Starburst.fd
\newcommand*\initfamily{\usefont{U}{Starburst}{xl}{n}}\initfamily

\makeatletter 
    
\renewcommand\onecolumngrid{
\do@columngrid{one}{\@ne}%
\def\set@footnotewidth{\onecolumngrid}
\def\footnoterule{\kern-6pt\hrule width 1.5in\kern6pt}%
}

\renewcommand\twocolumngrid{
        \def\footnoterule{
        \dimen@\skip\footins\divide\dimen@\thr@@
        \kern-\dimen@\hrule width.5in\kern\dimen@}
        \do@columngrid{mlt}{\tw@}
}%

\makeatother

\newcommand{\beq}{\begin{eqnarray}}
\newcommand{\eeq}{\end{eqnarray}}
\usepackage{amsmath}
\usepackage{tikz}
\usetikzlibrary{decorations.pathmorphing}
\usetikzlibrary{shapes.misc}
\tikzset{cross/.style={cross out, draw=black, minimum size=8*(#1-\pgflinewidth), inner sep=0pt, outer sep=0pt},
cross/.default={1pt}}
\usetikzlibrary{patterns,math}
\usepackage[colorlinks = true,
            linkcolor = blue,
            urlcolor  = blue,
            citecolor =red,
            anchorcolor = blue]{hyperref}

\begin{document}

\title{Thermalization dynamics of finite-size quantum critical systems
}

 \author{Li Li$^{1,2,3}$}
\email{liliphy@itp.ac.cn}
 \author{Yan Liu$^{4,5}$}
\email{yanliu@buaa.edu.cn}
 \author{Hao-Tian Sun$^{5}$}
 \email{sunhaotian@buaa.edu.cn}
 
\affiliation{
\vspace{0.25cm}
$^{1}$ Institute of Theoretical Physics,
Chinese Academy of Sciences, Beijing 100190, China}
\affiliation{$^2$School of Physical Sciences, University of Chinese Academy of Sciences, Beijing 100049, China}
\affiliation{$^3$School of Fundamental Physics and Mathematical Sciences, Hangzhou Institute for Advanced Study, UCAS, Hangzhou 310024, China}
\affiliation{$^4$Center for Gravitational Physics, Department of Space Science, 
Beihang University, Beijing 100191, China}
\affiliation{$^5$Peng Huanwu Collaborative Center for Research and Education, \\Beihang University, Beijing 100191, China}

\begin{abstract}
Using holographic duality, we investigate thermalization process when two finite-size quantum critical systems are brought into thermal contact along a perfectly transmitting interface. Through real-time simulations of gravitational dynamics, which are spatially inhomogeneous and anisotropic and are confined within two dynamical bulk branes, we identify three distinct thermalization patterns governed by the energy imbalance (temperature difference) and system size. For systems with large size and small energy imbalance, we observe recurrent cycles of formation and collapse of non-equilibrium steady states (NESS). Under large energy imbalance, shock waves persist for a prolonged period with sustained boundary reflections, while rarefaction waves rapidly homogenize. When the system size is sufficiently small, dissipation dominates and leads to oscillatory decay without sustained NESS or shock structure. In sharp contrast to  diffusive systems, we uncover that 
wave-propagated energy transfer  together with  boundary reflections enables nearly complete energy swapping between subsystems during thermalization. Our results reveal 
rich thermalization 
dynamics in finite-size quantum critical systems across  spatial scales and energy gradient regimes.
\end{abstract}

\maketitle

\textit{Introduction.--} 
Understanding far-from-equilibrium dynamics in quantum critical systems represents a frontier in modern physics, with profound implications across disciplines. These strongly correlated states underpin exotic phenomena in high-energy physics, condensed matter, and quantum information. A central paradigm involves thermalization of isolated system \cite{choi2016, delacretaz2025, kaufman2016, kinoshita2006, langen2015, le2023, neill2016, pulikkottil2023, rigol2009, shiraishi2024,gogolin2016,karrasch2013}.

\begin{figure*}[t]
		\begin{center}
		\begin{minipage}[b]{0.88\textwidth}
	    \includegraphics[width=1\textwidth]{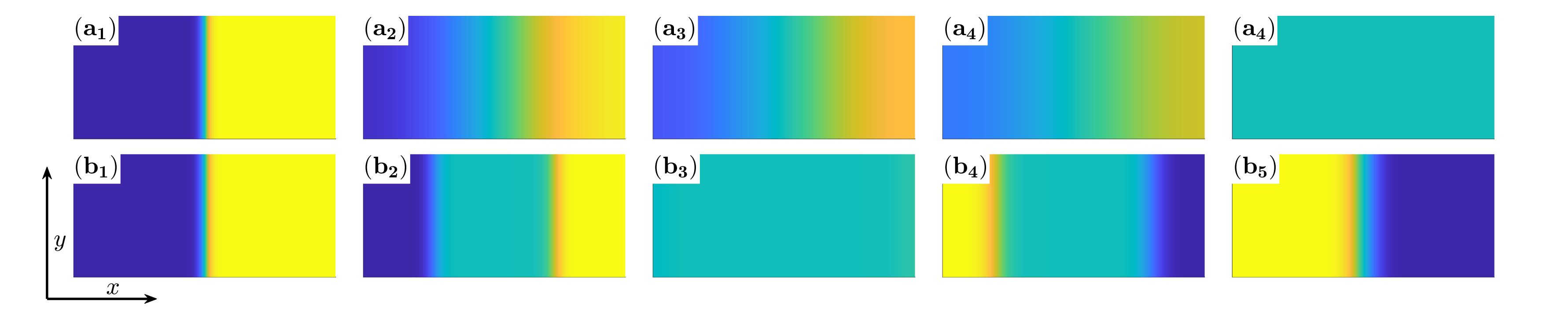}	
		\end{minipage}
		\begin{minipage}[b]{0.08\textwidth}
		      \includegraphics[width=1\textwidth]{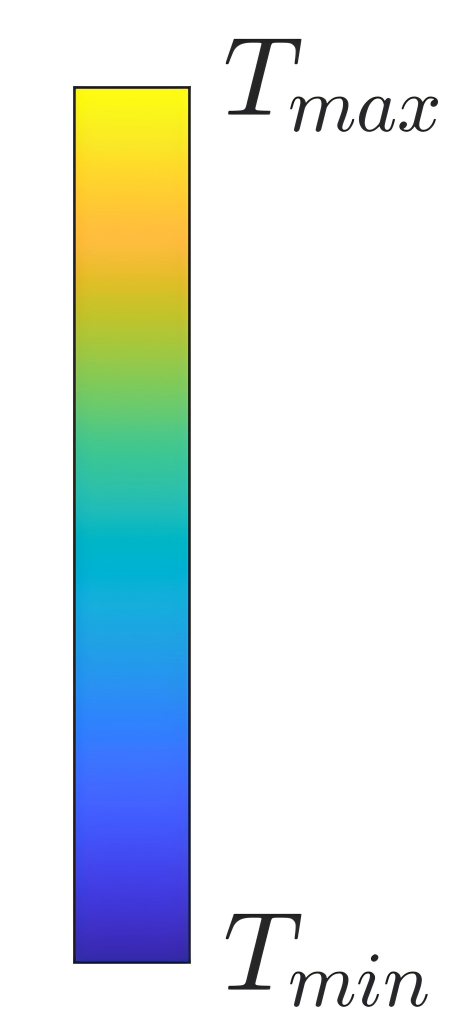}      
		\end{minipage}
		\includegraphics[width=0.65\textwidth]{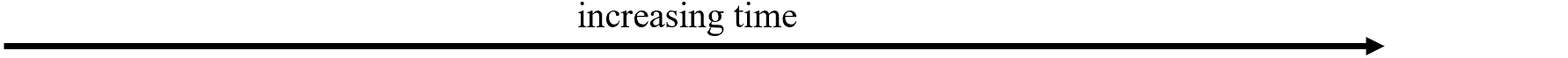}
			\caption{
		This schematic illustrates  thermalization process in two spatial dimensional ordinary matter ({\em top}) and a quantum critical system ({\em bottom}). At the beginning, two finite-size subsystems, independently thermalized at different temperatures, are brought into thermal contact along a perfectly transmitting interface ($a_1$ and $b_1$). Ordinary matter homogenizes as  temperature gradients diminish over time. The quantum critical system exhibits distinct thermalization dynamics -- one typical behavior we found in our work: a central NESS region forms ($b_2$), expands via shock and rarefaction waves, concentrate energy at the boundary ($b_4$), nearly mirrors the initial state ($b_5$) (with dissipation), then reverses through an expanding contact region, ultimately oscillating toward global equilibrium.}
			\label{fig:1}
		\end{center}
\end{figure*}

In conventional materials, thermal transport obeys Fourier's law, where the heat current proportional to the temperature gradients. By contrast, quantum critical systems governed by conformal symmetry exhibit profoundly distinct behavior --- as illustrated in Fig.~\ref{fig:1}. The top panels show evolution of thermalization  following Fourier's law, while the bottom panels display one representative result from our study for quantum critical system. When two such critical  systems at temperatures $T_L$, $T_R$
are brought into contact, they form a spatially homogeneous, non-equilibrium steady state (NESS) that sustains a persistent energy current without an energy density gradient \cite{bernard2012, bhaseen2015, lucas2016,chang2014}. In an infinitely extended system, pioneering work has revealed that this NESS corresponds to a Lorentz-boosted thermal distribution carrying a ballistic energy current sustained indefinitely. Through  holographic duality, this state is mapped  to a boosted black brane in anti-de Sitter (AdS) space~\cite{amado2015, bhaseen2015, ecker2021}.

However, real-world quantum systems, such as optical lattices and nanowires, are typically finite in size and subject to unavoidable dissipation.  These factors introduce boundary effects that fundamentally alter the system's  dynamics.  One crucial aspect is the thermalization dynamics, which  describes the evolution of the system from a non-equilibrium initial state to global thermal equilibrium.
Unlike in infinite systems,  NESS in finite-size quantum critical systems is expected to evolve toward global thermal equilibrium. 
These features pose critical challenges for energy transport, which becomes a transient process that highly sensitive to both boundary conditions and system size. First, boundary conditions can alter the thermalization dynamics. In an isolated system, for instance, the heat current must vanish at the boundary to conserve energy. In contrast, if the boundary temperature is fixed and non-uniform, a persistent internal heat current arises, preventing the system from reaching thermal equilibrium. Second, the finite system size introduces an intrinsic scale that influences gradients of dynamical quantities, resulting in distinct evolutionary pathways during thermalization. 

These complexities make traditional field theory approaches intractable---especially in systems with small size, 
strong nonlinearities, and strong coupling---which can lead to the breakdown of hydrodynamics and linear-response approximations. The AdS/CFT correspondence provides a powerful framework by encoding quantum dynamics in a gravitational spacetime geometry~\cite{chesler2014, hubeny2010, liu2018, cardoso2012}.  
Its extension, AdS/BCFT~\cite{fujita2011, nozaki2012, takayanagi2011}, incorporates boundaries into the CFT through an ``end-of-the-world" (EOW) brane in the AdS bulk, anchored to the CFT boundary. This brane modifies the bulk geometry and encodes the boundary conditions of the CFT. The framework provides geometric insights into boundary phenomena, such as entanglement entropy near edges, defects, and impurity effects, with broad  applications in quantum gravity, condensed matter physics, and strongly coupled systems~\cite{Andrei:2018die,izumi2022}. 
AdS/BCFT provides 
a powerful tool for studying holographic dualities in constrained geometries and enabling investigations of how far-from-equilibrium finite systems evolve toward global thermal equilibrium. In this letter, we investigate the thermalization dynamics of an isolated, non-equilibrium finite-size system by considering the full dynamics of bulk spacetime.
\\

\textit{Holographic setup of finite size system.--} 
The real time evolution dynamics of strongly couple system can be studied using a holographic model in asymptotically AdS$_4$ spacetime, with an action 
\begin{equation}\label{action}
S\,=\frac{1}{2\kappa_N^2}\,\int d^4x \sqrt{-g}
\left[\mathcal{R}-2 \Lambda\right]\,,
\end{equation}
with $\mathcal{R}$ the Ricci scalar, $\Lambda$ the cosmological constant, and $\kappa_N^2=8\pi G_N$ related to Newton’s constant. The dual field theory lives in $2+1$ dimension. We set $\kappa_N^2=1$ and $\Lambda=-3$ for simplicity.

To incorporate boundary effects, we introduce dynamical boundaries (EOW brane $Q$) in bulk asymptotically AdS spacetime (the spatial slice is shown in Fig. S1 in
Appendix~\ref{holographic setup})
, which is dual to a BCFT defined on a manifold $\mathcal{M}$ with boundary $\partial\mathcal{M}$. We consider a strip shape boundary system where one spatial direction is confined by $Q$ and the other direction is extended homogeneous. The Gibbons-Hawking term is included to ensure a well-posed variational principle. Following the AdS/BCFT framework, we impose Neumann boundary conditions on $Q$, 
\begin{align}
    K_{\mu\nu}-K\gamma_{\mu\nu}=\hat{T}_{\mu\nu}
\end{align}
where $\gamma_{\mu\nu}$ is the induced metric on $Q$,  $K_{\mu\nu}=\gamma_{\mu}{}^\rho\nabla_{\rho}n_{\nu}$ the extrinsic curvature of $Q$, $n_\nu$ the unit vector normal to $Q$,  $K=\gamma^{\mu\nu}K_{\mu\nu}$, and $\hat{T}_{\mu\nu}$ the energy-momentum tensor of brane-localized, which governs  boundary degrees of freedom in the dual BCFT. Here, we choose $\hat{T}_{\mu\nu}=0$, {\em i.e.} a tensionless brane without matter. 

To model an isolated system, we 
impose the following boundary condition on $\partial\mathcal{M}$ for the energy-momentum tensor of the dual field theory 
\begin{align}\label{current}
    T^{tx}\equiv J=0,\quad x\in\partial\mathcal{M}\,.
\end{align}
This condition, along with $\partial_a T^{ab}=0$, ensures the total energy conservation in the isolated system.

We study the thermalization process by preparing a far-from-equilibrium initial state considering two subsystems with different temperatures brought into contact at the center (as shown in Fig.~\ref{fig:1}). Since the definition of temperature is ambiguous when the system is far-from equilibrium, we use energy density --  directly accessible via  holographic dictionary -- as the relevant observable. We choose the initial energy density profile to be
\begin{align}
    E(x)=E_L+(E_R-E_L)\left(1+\tanh\frac{x}{\alpha L}\right)\label{E_ini}
\end{align}
where $x\in\left[-L,L\right]$ is the  spacial coordinate, and $\alpha$ a dimensionless parameter controlling the sharpness of the interface. For sufficiently small $\alpha$ (here we choose $\alpha=0.07$), the profile approaches a step function without qualitatively altering the dynamics. 
The left and right regions have energy densities $E_L$ and $E_R$. This results in a system involving ten cohomogeneity-three PDE’s from the Einstein  equations, together with the challenge for consistently dealing with the bulk brane $Q$. By numerically solving the system, we observe nonlinear dynamics evolving toward global thermal equilibrium. Further details on the equations of motion and numerical time evolution can be found in Appendix~\ref{holographic setup}, ~\ref{time evolution}  and \ref{erranalysis}. 

\begin{figure}[h]
	 \centering
	  \includegraphics[width=0.48\textwidth]{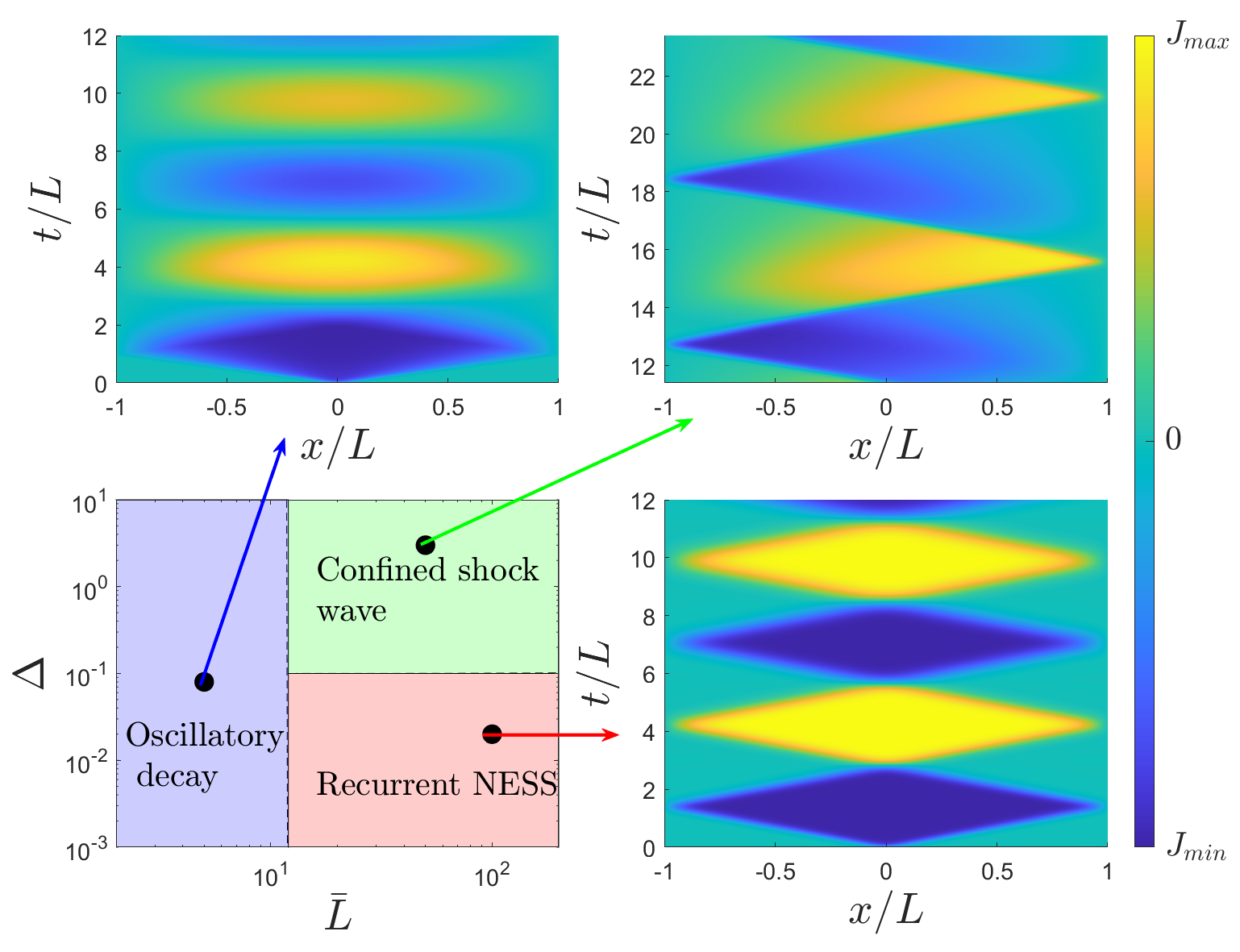}
	    \caption{Numerical solutions for the finite size quantum critical system. Density plots show the space-time evolution of energy energy current $J$ for different  energy imbalance $\Delta$ and system size $L$, including  $\Delta=0.08,\ \bar{L}=5$ for pure oscillatory decay ({\em top-left}), $\Delta=3,\ \bar{L}=50$ for confined shock wave ({\em top-right}), and $\Delta=3,\ \bar{L}=50$ for recurrent NESS ({\em bottom-right}). The phase diagram is shown in the  {\em bottom-left panel}, where the demarcation between different regions is schematic since blending of distinct characteristics occurs near the boundary.
	    }
	    \label{fig:NESS_Ttx}
\end{figure}

During this process, the dynamics are governed by two dimensionless quantities:  the relative energy difference between the left and right systems,  $\Delta=\left(E_R-E_L\right)/E_L$, and the rescaled system size  $\Bar{L}=L\left(E_R+E_L\right)^{1/3}/2^{1/3}$. As illustrated in Fig.~\ref{fig:NESS_Ttx}, varying $\Delta$ and $\Bar{L}$ leads to three distinct dynamical behaviors before settling down to global equilibrium.
\\

\textit{Recurrent NESS.--}
When $\Bar{L}$ is large and $\Delta$ is small,  a NESS region initially forms in the contact region, as shown in the bottom panel of  Fig.~\ref{fig:1}. This is consistent with previous studies \cite{amado2015, bhaseen2015, ecker2021}, since the boundaries are not yet influencing the dynamics. The NESS region expands via a shock wave (with higher energy density behind the wavefront) and a rarefaction  wave (with higher energy density ahead of the wavefront). Both the shock wave velocity $v_s$ and the rarefaction wave velocity $v_r$ are approximately equal to the speed of sound wave $c_s=1/\sqrt{2}$.

\begin{figure}[h]
	 
	 \begin{minipage}{0.48\textwidth}
	 
	  	  \includegraphics[width=1\textwidth]{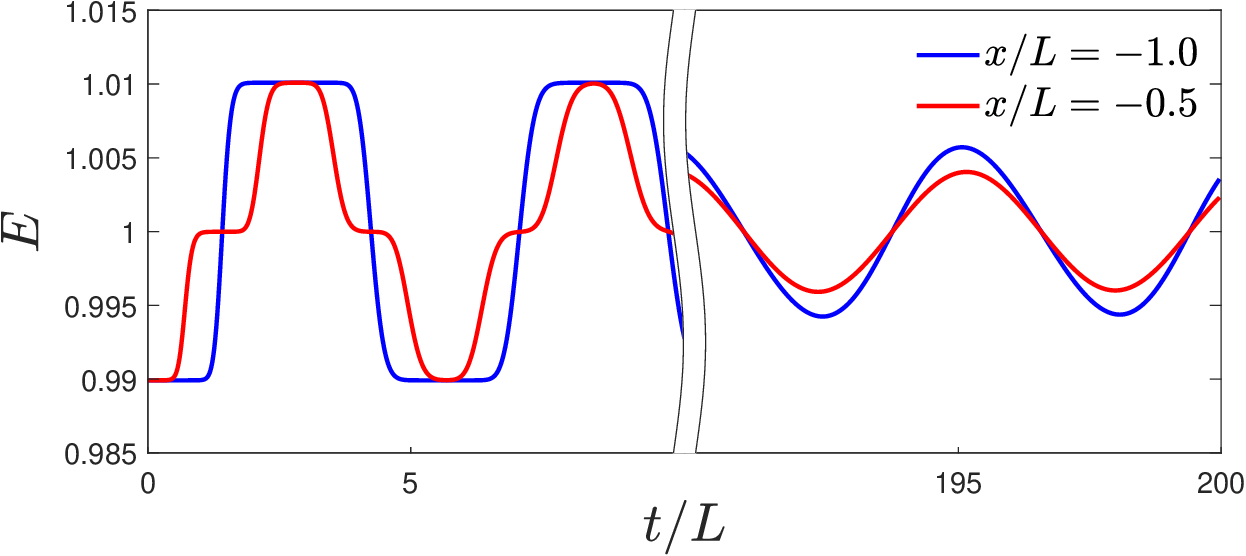}
	 \end{minipage}
	 \hfill
    \begin{minipage}{0.47\textwidth}
    \includegraphics[width=1\textwidth]{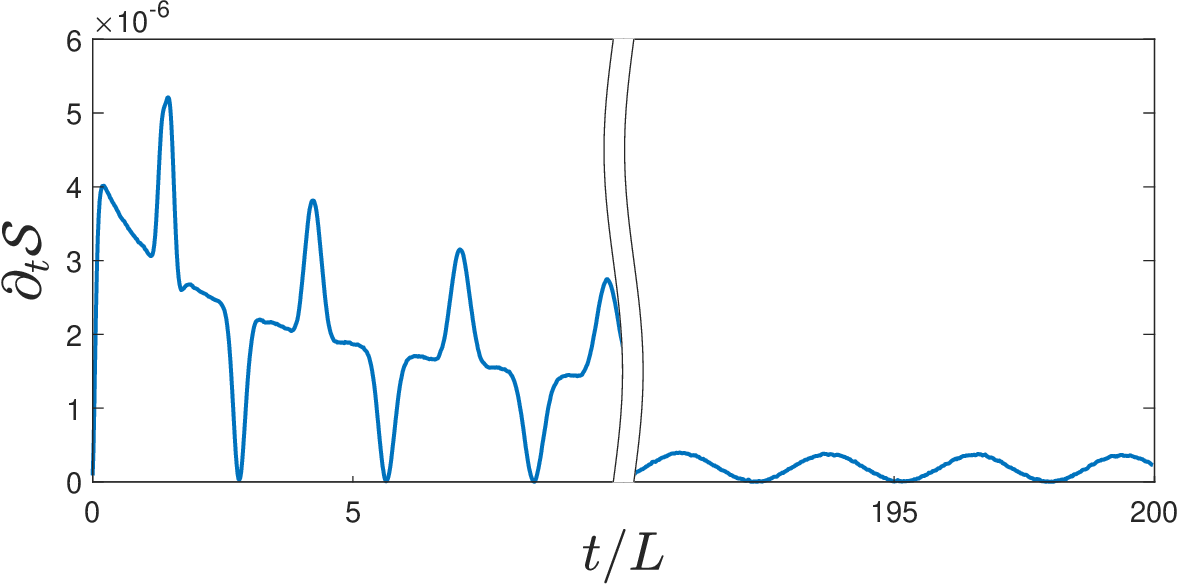}
	 \end{minipage}
	  
	    \caption{Far-from-equilibrium dynamics for $\Delta=0.02,\ \Bar{L}=100$. The system change from the  recurrent NESS stage to a phase of oscillatory decay at late times. {\em Top panel:} Time evolution of the  energy density at two spatial points $x/L=-1$ (blue) and $x/L=-0.5$ (red). {\em Bottom panel:} Time evolution of the time derivative of the total entropy.  }
	    \label{fig:NESS_energy_entropy_t}
\end{figure}

Since the system size is finite, as time evolves, the NESS region inevitably reaches the boundary. In an isolated system described by Eq.~\eqref{current}, it then reflects back. The energy current $J$, shown in the bottom-right panel of Fig.~\ref{fig:NESS_Ttx}, undergoes ballistic changes across spatial locations during this process. 
Both the energy density and energy current at a given point $x_0$ vary only when the shock wave or rarefaction wave passes through it, as illustrated in the top panel of Fig.~\ref{fig:NESS_energy_entropy_t}. Approximately after a time $t=\frac{2L}{c_s}$, the system will returns to a nearly mirror reflection state, the energy current is zero but the energy density is mirror reflected (as shown in bottom panel of Fig.~\ref{fig:1}). In the profile of energy flow, a series of diamond zones are formed (bottom-right panel of Fig.~\ref{fig:NESS_Ttx}).

After the  recurrent NESS stage, the system enters into a phase of oscillatory decay. Rather than exhibiting random behavior, a synchronized pattern emerges. At late times, the energy density follows the form
\begin{align}
    E\propto e^{-t/\tau}\cos(\omega t)\sin\frac{\pi x}{2L}\,,
\end{align}
where the constant $\tau$ characterises the dissipation rate and $\omega$ denotes the oscillation  frequency. The entire dynamical process -- from the recurrent NESS to oscillatory decay -- is illustrated in the top panel of Fig.~\ref{fig:NESS_energy_entropy_t}.

While the total energy of the system is conserved, there is entropy production in the system. Entropy production related to the gradient of fluid velocity. We 
compute the total entropy of the boundary system, $\mathcal{S}=\mathcal{A}/4G_N$, 
using the area of apparent horizon~\cite{engelhardt2018,Baggioli:2021tzr, hollands2024, rougemont2022}. The total entropy production can thus be extracted from the increase in the apparent horizon area, which is shown in the bottom panel of Fig.~\ref{fig:NESS_energy_entropy_t} (see also Appendix \ref{energy conservation}). A clear correlation between the coevolution of energy density and entropy production is evident. Specifically, the entropy production peaks when the NESS reaches the boundary,  and reaches zero when the energy current vanishes (see bottom-right panel in Fig.~\ref{fig:NESS_Ttx}).
\\

\textit{Confined shock wave.--}
As $\Delta$ increases, the shock wave and rarefaction wave exhibit distinct  propagation behaviors. As illustrated in the top-right panel of Fig.~\ref{fig:NESS_Ttx}, the rarefaction wave broadens more rapidly than the shock wave, quickly propagating throughout the system and subsequently smoothing out. In contrast, the shock wave persists for a considerably longer duration and evolves into a confined structure which is different from the standard shock wave that is approximately localized at a fixed position in an infinity size system. In our case, the shock wave is always affected by the boundary, therefore we term it as ``confined shock wave". As a result, no clear formation of a NESS is observed. 

\begin{figure}[h]
	 \centering
	  \includegraphics[width=0.48\textwidth]{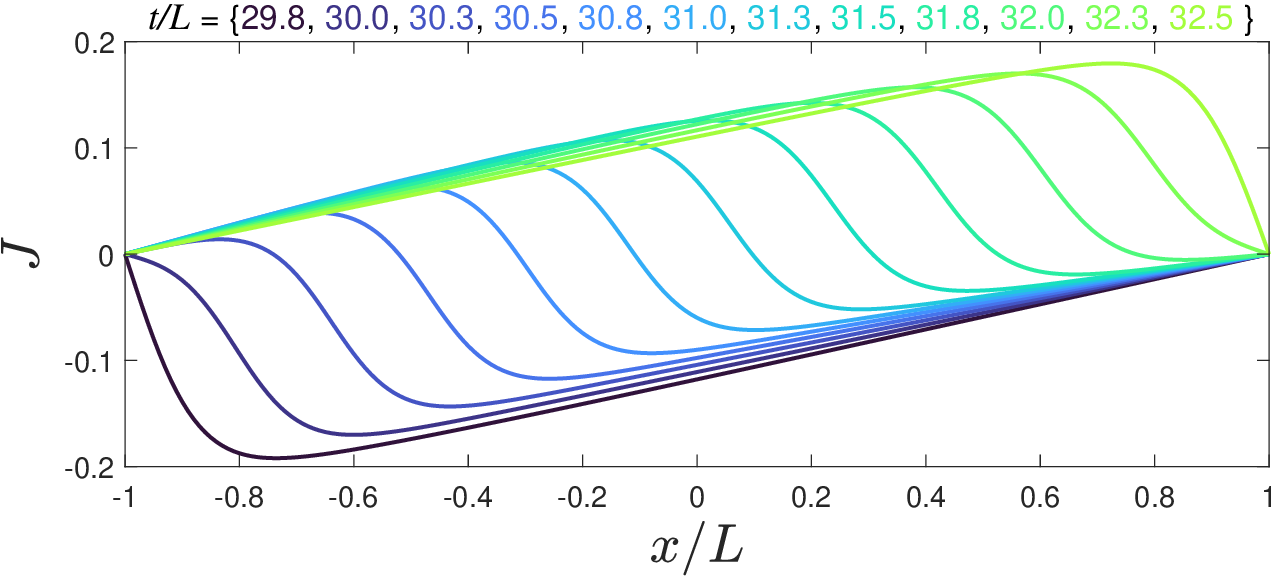}
	  \includegraphics[width=0.48\textwidth]{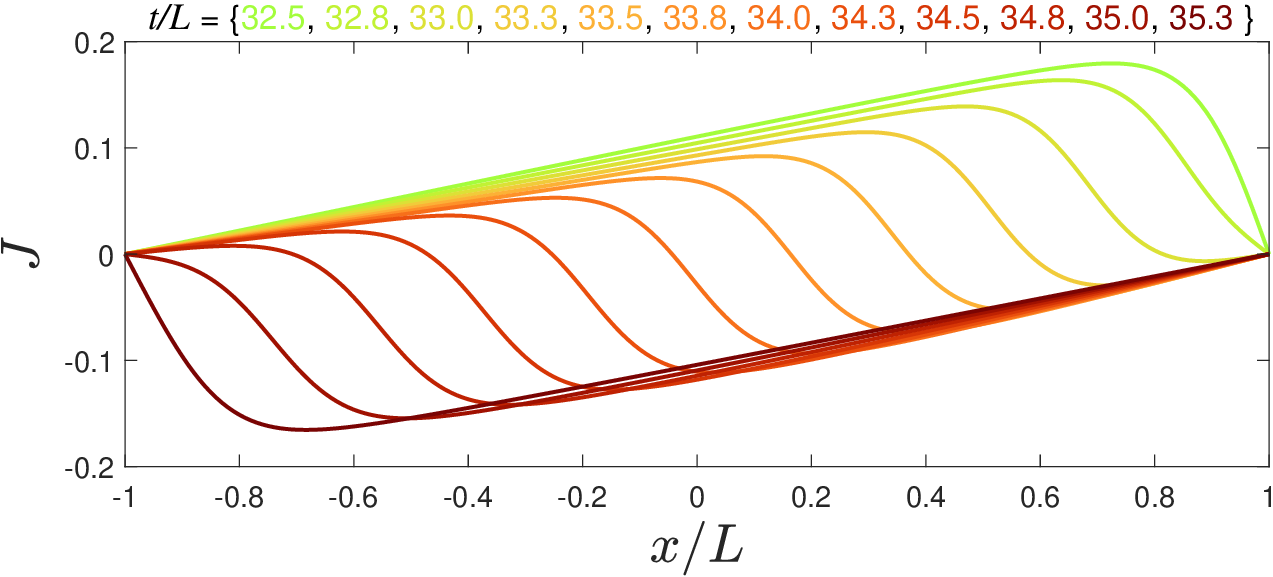}
	    \caption{Profiles of the energy current $J$  at different times for $\Delta=3,\ \Bar{L}=50$. The shock wave reflects repeatedly between the boundaries. Note that $J=0$ at both boundaries $x=\pm L$, consistent with an isolated system with no energy enters or exits through the boundaries.}
	    \label{fig:shock_energy_current}
\end{figure}

After the rarefaction wave has smoothed out, the time evolution of the energy current --- shown in Fig.~\ref{fig:shock_energy_current} --- reveals a persistent spatial gradient.  Consequently, unlike the periodic recurrence of 
zero entropy production observed in the bottom panel of  Fig.~\ref{fig:NESS_energy_entropy_t},  entropy production at this stage no longer vanishes periodically. Nevertheless, a distinct peak remains visible each time the shock wave reaches the boundary (middle panel of Fig.~S2). Similarly, dissipation leads to the broadening of the shock wave and causes it to eventually decay. Subsequently, the entire system synchronizes and undergoes collective decay, similar to what is shown in Fig.~\ref{fig:NESS_energy_entropy_t}.

~\\

\textit{Oscillatory decay.--}
As the system size continues to decrease, we observe that  dissipation becomes the dominate factor governing the dynamics. As a result, neither  shock wave nor NESS emerges emerge; instead, the system synchronizes and rapidly enters a phase of oscillatory decay. Moreover, as  increasing $\Delta$ increases, higher modes are found to be excited, but no persistent structure forms. The energy current corresponding to this type of dynamics—for a relative small $\Delta$—is shown in the bottom-right panel of Fig.~\ref{fig:NESS_Ttx}.

\begin{figure}[h]
	 \centering
	  \includegraphics[width=0.46\textwidth]{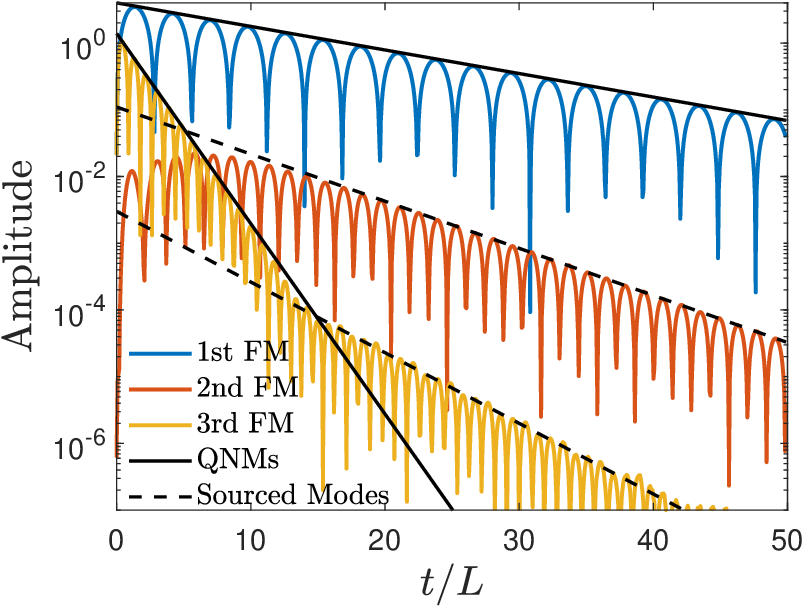}
	    \caption{ Temporal evolution of the FM amplitudes of energy current for $\Delta=0.08,\ \Bar{L}=5$. Solid black lines comes from the linear QNMs around the final equilibrium black hole, while the dashed lines represent nonlinear modes originating  from higher-order perturbations of the final black hole spacetime.}
	    \label{fig:qnm_t}
\end{figure}

To quantitatively understand the system's  behavior, we extract the evolution of each Fourier mode (FM) by preforming an odd extension of the energy current, see Fig.~\ref{fig:qnm_t}. Initially, the configuration excites all even-$n$ modes. Due to nonlinear effects, the amplitude of the second FM increases, resulting in  energy transfer across different wavelengths. Nevertheless, the first and third FMs are well captured by the quasi-normal modes (QNMs) derived from linear black hole perturbation theory around the final equilibrium black hole state.

Notably, the decay rate of the third mode differs between the early and late stages. This occurs because the first mode --- which is substantially larger than higher-order modes --- acts as a source term in their perturbation equations, ultimately driving the system into a forced oscillatory regime. A similar phenomenon, known as nonlinear QNMs, has also been observed in gravitational wave  detection  \cite{baibhav2023, cheung2023, cheung2024}. A detailed analysis can be found in Appendix~\ref{pertanalusis}. 
\\

\textit{Conclusions and discussions.--} Combining holography and numerical general relativity, we have simulated a real time dynamics of thermalization in strongly coupled quantum systems of finite size with dissipation. Our results reveal that the system size $L$ has a strikingly pronounced effect on the non-equilibrium dynamics. for large $L$
boundaries reflect energy currents back into the system. Moreover, due to the competition between the shock wave and the rarefaction wave, which is controlled by the energy difference 
$\Delta$, we observe recurrent NESS for small $\Delta$ and long-lived confined shock waves for large $\Delta$. In contrast, For small $L$, internal dissipation dominates and disrupts any structures that would otherwise form in  larger system sizes. A schematic phase diagram is presented in the bottom-left panel of Fig.~\ref{fig:NESS_Ttx}.

We have observed a distinct difference in energy transfer between quantum critical system and conventional matter. As shown in the bottom panel of Fig.~\ref{fig:1}, if the two systems are separated after contact, the temperatures of the left and right systems undergo an approximate direct swap. This behavior does not violate the second law of thermodynamics, since the total entropy increases (Fig.~\ref{fig:NESS_Ttx}). The underlying mechanism involves wave-like energy propagation and reflection 
due to the finite-size system's boundary. 
This phenomenon may potentially be harnessed for cooling applications. 

~\\

While our analysis has been limited to the two spatial dimensions, 
the generalization to other dimensions is straightforward. Moreover, this approach enables the construction of a holographic description for open finite-size systems through modifications of boundary conditions on the EOW brane.
Finally, it is interesting to include other matter fields in the bulk to study, \emph{e.g.} the charge transport for finite-size systems. 

\subsection*{Acknowledgments} 
We thank Elias Kiritsis, Wilke van der Schee and Javier G. Subils for helpful discussions. This work was supported by the National Natural Science Foundation of China Grants No. 12525503, No. 12375041, No. 12447169, No. 12575046 and No. 12447101. H-T. S also acknowledges the support from the Postdoctoral Fellowship Program of CPSF under Grant No. GZC20252777. We acknowledge the use of the High Performance Cluster at Institute of Theoretical Physics, Chinese Academy of Sciences.

\bibliographystyle{modified-apsrev4-2.bst}
\bibliography{ref}

\begin{thebibliography}{34}%
\makeatletter
\providecommand \@ifxundefined [1]{%
 \@ifx{#1\undefined}
}%
\providecommand \@ifnum [1]{%
 \ifnum #1\expandafter \@firstoftwo
 \else \expandafter \@secondoftwo
 \fi
}%
\providecommand \@ifx [1]{%
 \ifx #1\expandafter \@firstoftwo
 \else \expandafter \@secondoftwo
 \fi
}%
\providecommand \natexlab [1]{#1}%
\providecommand \enquote  [1]{``#1''}%
\providecommand \bibnamefont  [1]{#1}%
\providecommand \bibfnamefont [1]{#1}%
\providecommand \citenamefont [1]{#1}%
\providecommand \href@noop [0]{\@secondoftwo}%
\providecommand \href [0]{\begingroup \@sanitize@url \@href}%
\providecommand \@href[1]{\@@startlink{#1}\@@href}%
\providecommand \@@href[1]{\endgroup#1\@@endlink}%
\providecommand \@sanitize@url [0]{\catcode `\\12\catcode `\$12\catcode
  `\&12\catcode `\#12\catcode `\^12\catcode `\_12\catcode `\%12\relax}%
\providecommand \@@startlink[1]{}%
\providecommand \@@endlink[0]{}%
\providecommand \url  [0]{\begingroup\@sanitize@url \@url }%
\providecommand \@url [1]{\endgroup\@href {#1}{\urlprefix }}%
\providecommand \urlprefix  [0]{URL }%
\providecommand \Eprint [0]{\href }%
\providecommand \doibase [0]{https://doi.org/}%
\providecommand \selectlanguage [0]{\@gobble}%
\providecommand \bibinfo  [0]{\@secondoftwo}%
\providecommand \bibfield  [0]{\@secondoftwo}%
\providecommand \translation [1]{[#1]}%
\providecommand \BibitemOpen [0]{}%
\providecommand \bibitemStop [0]{}%
\providecommand \bibitemNoStop [0]{.\EOS\space}%
\providecommand \EOS [0]{\spacefactor3000\relax}%
\providecommand \BibitemShut  [1]{\csname bibitem#1\endcsname}%
\let\auto@bib@innerbib\@empty
\bibitem [{\citenamefont {Choi}\ \emph {et~al.}(2016)\citenamefont {Choi},
  \citenamefont {Hild}, \citenamefont {Zeiher}, \citenamefont {Schau{\ss}},
  \citenamefont {{Rubio-Abadal}}, \citenamefont {Yefsah}, \citenamefont
  {Khemani}, \citenamefont {Huse}, \citenamefont {Bloch},\ and\ \citenamefont
  {Gross}}]{choi2016}%
  \BibitemOpen
  \bibfield  {author} {\bibinfo {author} {\bibfnamefont {J.-y.}\ \bibnamefont
  {Choi}}, \bibinfo {author} {\bibfnamefont {S.}~\bibnamefont {Hild}}, \bibinfo
  {author} {\bibfnamefont {J.}~\bibnamefont {Zeiher}}, \bibinfo {author}
  {\bibfnamefont {P.}~\bibnamefont {Schau{\ss}}}, \bibinfo {author}
  {\bibfnamefont {A.}~\bibnamefont {{Rubio-Abadal}}}, \bibinfo {author}
  {\bibfnamefont {T.}~\bibnamefont {Yefsah}}, \bibinfo {author} {\bibfnamefont
  {V.}~\bibnamefont {Khemani}}, \bibinfo {author} {\bibfnamefont {D.~A.}\
  \bibnamefont {Huse}}, \bibinfo {author} {\bibfnamefont {I.}~\bibnamefont
  {Bloch}},\ and\ \bibinfo {author} {\bibfnamefont {C.}~\bibnamefont {Gross}},\
  }\bibfield  {title} {\bibinfo {title} {{\color{black}Exploring the Many-Body
  Localization Transition in Two Dimensions}},\ }\href
  {https://doi.org/10.1126/science.aaf8834} {\bibfield  {journal} {\bibinfo
  {journal} {Science}\ }\textbf {\bibinfo {volume} {352}},\ \bibinfo {pages}
  {1547} (\bibinfo {year} {2016})}\BibitemShut {NoStop}%
\bibitem [{\citenamefont {Delacr{\'e}taz}(2025)}]{delacretaz2025}%
  \BibitemOpen
  \bibfield  {author} {\bibinfo {author} {\bibfnamefont {L.~V.}\ \bibnamefont
  {Delacr{\'e}taz}},\ }\bibfield  {title} {\bibinfo {title} {{\color{black}A
  Bound on Thermalization from Diffusive Fluctuations}},\ }\href
  {https://doi.org/10.1038/s41567-024-02774-9} {\bibfield  {journal} {\bibinfo
  {journal} {Nature Physics}\ ,\ \bibinfo {pages} {1}} (\bibinfo {year}
  {2025})}\BibitemShut {NoStop}%
\bibitem [{\citenamefont {Kaufman}\ \emph {et~al.}(2016)\citenamefont
  {Kaufman}, \citenamefont {Tai}, \citenamefont {Lukin}, \citenamefont
  {Rispoli}, \citenamefont {Schittko}, \citenamefont {Preiss},\ and\
  \citenamefont {Greiner}}]{kaufman2016}%
  \BibitemOpen
  \bibfield  {author} {\bibinfo {author} {\bibfnamefont {A.~M.}\ \bibnamefont
  {Kaufman}}, \bibinfo {author} {\bibfnamefont {M.~E.}\ \bibnamefont {Tai}},
  \bibinfo {author} {\bibfnamefont {A.}~\bibnamefont {Lukin}}, \bibinfo
  {author} {\bibfnamefont {M.}~\bibnamefont {Rispoli}}, \bibinfo {author}
  {\bibfnamefont {R.}~\bibnamefont {Schittko}}, \bibinfo {author}
  {\bibfnamefont {P.~M.}\ \bibnamefont {Preiss}},\ and\ \bibinfo {author}
  {\bibfnamefont {M.}~\bibnamefont {Greiner}},\ }\bibfield  {title} {\bibinfo
  {title} {{\color{black}Quantum Thermalization through Entanglement in an
  Isolated Many-Body System}},\ }\href
  {https://doi.org/10.1126/science.aaf6725} {\bibfield  {journal} {\bibinfo
  {journal} {Science}\ }\textbf {\bibinfo {volume} {353}},\ \bibinfo {pages}
  {794} (\bibinfo {year} {2016})}\BibitemShut {NoStop}%
\bibitem [{\citenamefont {Kinoshita}\ \emph {et~al.}(2006)\citenamefont
  {Kinoshita}, \citenamefont {Wenger},\ and\ \citenamefont
  {Weiss}}]{kinoshita2006}%
  \BibitemOpen
  \bibfield  {author} {\bibinfo {author} {\bibfnamefont {T.}~\bibnamefont
  {Kinoshita}}, \bibinfo {author} {\bibfnamefont {T.}~\bibnamefont {Wenger}},\
  and\ \bibinfo {author} {\bibfnamefont {D.~S.}\ \bibnamefont {Weiss}},\
  }\bibfield  {title} {\bibinfo {title} {{\color{black}A Quantum {{Newton}}'s
  Cradle}},\ }\href {https://doi.org/10.1038/nature04693} {\bibfield  {journal}
  {\bibinfo  {journal} {Nature}\ }\textbf {\bibinfo {volume} {440}},\ \bibinfo
  {pages} {900} (\bibinfo {year} {2006})}\BibitemShut {NoStop}%
\bibitem [{\citenamefont {Langen}\ \emph {et~al.}(2015)\citenamefont {Langen},
  \citenamefont {Erne}, \citenamefont {Geiger}, \citenamefont {Rauer},
  \citenamefont {Schweigler}, \citenamefont {Kuhnert}, \citenamefont
  {Rohringer}, \citenamefont {Mazets}, \citenamefont {Gasenzer},\ and\
  \citenamefont {Schmiedmayer}}]{langen2015}%
  \BibitemOpen
  \bibfield  {author} {\bibinfo {author} {\bibfnamefont {T.}~\bibnamefont
  {Langen}}, \bibinfo {author} {\bibfnamefont {S.}~\bibnamefont {Erne}},
  \bibinfo {author} {\bibfnamefont {R.}~\bibnamefont {Geiger}}, \bibinfo
  {author} {\bibfnamefont {B.}~\bibnamefont {Rauer}}, \bibinfo {author}
  {\bibfnamefont {T.}~\bibnamefont {Schweigler}}, \bibinfo {author}
  {\bibfnamefont {M.}~\bibnamefont {Kuhnert}}, \bibinfo {author} {\bibfnamefont
  {W.}~\bibnamefont {Rohringer}}, \bibinfo {author} {\bibfnamefont {I.~E.}\
  \bibnamefont {Mazets}}, \bibinfo {author} {\bibfnamefont {T.}~\bibnamefont
  {Gasenzer}},\ and\ \bibinfo {author} {\bibfnamefont {J.}~\bibnamefont
  {Schmiedmayer}},\ }\bibfield  {title} {\bibinfo {title}
  {{\color{black}Experimental Observation of a Generalized {{Gibbs}}
  Ensemble}},\ }\href {https://doi.org/10.1126/science.1257026} {\bibfield
  {journal} {\bibinfo  {journal} {Science}\ }\textbf {\bibinfo {volume}
  {348}},\ \bibinfo {pages} {207} (\bibinfo {year} {2015})}\BibitemShut
  {NoStop}%
\bibitem [{\citenamefont {Le}\ \emph {et~al.}(2023)\citenamefont {Le},
  \citenamefont {Zhang}, \citenamefont {Gopalakrishnan}, \citenamefont
  {Rigol},\ and\ \citenamefont {Weiss}}]{le2023}%
  \BibitemOpen
  \bibfield  {author} {\bibinfo {author} {\bibfnamefont {Y.}~\bibnamefont
  {Le}}, \bibinfo {author} {\bibfnamefont {Y.}~\bibnamefont {Zhang}}, \bibinfo
  {author} {\bibfnamefont {S.}~\bibnamefont {Gopalakrishnan}}, \bibinfo
  {author} {\bibfnamefont {M.}~\bibnamefont {Rigol}},\ and\ \bibinfo {author}
  {\bibfnamefont {D.~S.}\ \bibnamefont {Weiss}},\ }\bibfield  {title} {\bibinfo
  {title} {{\color{black}Observation of Hydrodynamization and Local
  Prethermalization in {{1D Bose}} Gases}},\ }\href
  {https://doi.org/10.1038/s41586-023-05979-9} {\bibfield  {journal} {\bibinfo
  {journal} {Nature}\ }\textbf {\bibinfo {volume} {618}},\ \bibinfo {pages}
  {494} (\bibinfo {year} {2023})}\BibitemShut {NoStop}%
\bibitem [{\citenamefont {Neill}\ \emph {et~al.}(2016)\citenamefont {Neill}
  \emph {et~al.}}]{neill2016}%
  \BibitemOpen
  \bibfield  {author} {\bibinfo {author} {\bibfnamefont {C.}~\bibnamefont
  {Neill}} \emph {et~al.},\ }\bibfield  {title} {\bibinfo {title}
  {{\color{black}Ergodic Dynamics and Thermalization in an Isolated Quantum
  System}},\ }\href {https://doi.org/10.1038/nphys3830} {\bibfield  {journal}
  {\bibinfo  {journal} {Nature Phys.}\ }\textbf {\bibinfo {volume} {12}},\
  \bibinfo {pages} {1037} (\bibinfo {year} {2016})}\BibitemShut {NoStop}%
\bibitem [{\citenamefont {Pulikkottil}\ \emph {et~al.}(2023)\citenamefont
  {Pulikkottil}, \citenamefont {Lakshminarayan}, \citenamefont {Srivastava},
  \citenamefont {Kieler}, \citenamefont {B{\"a}cker},\ and\ \citenamefont
  {Tomsovic}}]{pulikkottil2023}%
  \BibitemOpen
  \bibfield  {author} {\bibinfo {author} {\bibfnamefont {J.~J.}\ \bibnamefont
  {Pulikkottil}}, \bibinfo {author} {\bibfnamefont {A.}~\bibnamefont
  {Lakshminarayan}}, \bibinfo {author} {\bibfnamefont {S.~C.~L.}\ \bibnamefont
  {Srivastava}}, \bibinfo {author} {\bibfnamefont {M.~F.~I.}\ \bibnamefont
  {Kieler}}, \bibinfo {author} {\bibfnamefont {A.}~\bibnamefont {B{\"a}cker}},\
  and\ \bibinfo {author} {\bibfnamefont {S.}~\bibnamefont {Tomsovic}},\
  }\bibfield  {title} {\bibinfo {title} {{\color{black}Quantum Coherence
  Controls the Nature of Equilibration and Thermalization in Coupled Chaotic
  Systems}},\ }\href {https://doi.org/10.1103/PhysRevE.107.024124} {\bibfield
  {journal} {\bibinfo  {journal} {Phys. Rev. E}\ }\textbf {\bibinfo {volume}
  {107}},\ \bibinfo {pages} {024124} (\bibinfo {year} {2023})}\BibitemShut
  {NoStop}%
\bibitem [{\citenamefont {Rigol}(2009)}]{rigol2009}%
  \BibitemOpen
  \bibfield  {author} {\bibinfo {author} {\bibfnamefont {M.}~\bibnamefont
  {Rigol}},\ }\bibfield  {title} {\bibinfo {title} {{\color{black}Breakdown of
  {{Thermalization}} in {{Finite One-Dimensional Systems}}}},\ }\href
  {https://doi.org/10.1103/PhysRevLett.103.100403} {\bibfield  {journal}
  {\bibinfo  {journal} {Physical Review Letters}\ }\textbf {\bibinfo {volume}
  {103}},\ \bibinfo {pages} {100403} (\bibinfo {year} {2009})}\BibitemShut
  {NoStop}%
\bibitem [{\citenamefont {Shiraishi}\ and\ \citenamefont
  {Tasaki}(2024)}]{shiraishi2024}%
  \BibitemOpen
  \bibfield  {author} {\bibinfo {author} {\bibfnamefont {N.}~\bibnamefont
  {Shiraishi}}\ and\ \bibinfo {author} {\bibfnamefont {H.}~\bibnamefont
  {Tasaki}},\ }\bibfield  {title} {\bibinfo {title} {{\color{black}Nature
  {{Abhors}} a {{Vacuum}}: {{A Simple Rigorous Example}} of {{Thermalization}}
  in an {{Isolated Macroscopic Quantum System}}}},\ }\href
  {https://doi.org/10.1007/s10955-024-03289-6} {\bibfield  {journal} {\bibinfo
  {journal} {J. Statist. Phys.}\ }\textbf {\bibinfo {volume} {191}},\ \bibinfo
  {pages} {82} (\bibinfo {year} {2024})}\BibitemShut {NoStop}%
\bibitem [{\citenamefont {Gogolin}\ and\ \citenamefont
  {Eisert}(2016)}]{gogolin2016}%
  \BibitemOpen
  \bibfield  {author} {\bibinfo {author} {\bibfnamefont {C.}~\bibnamefont
  {Gogolin}}\ and\ \bibinfo {author} {\bibfnamefont {J.}~\bibnamefont
  {Eisert}},\ }\bibfield  {title} {\bibinfo {title}
  {{\color{black}Equilibration, Thermalisation, and the Emergence of Statistical
  Mechanics in Closed Quantum Systems}},\ }\href
  {https://doi.org/10.1088/0034-4885/79/5/056001} {\bibfield  {journal}
  {\bibinfo  {journal} {Rept. Prog. Phys.}\ }\textbf {\bibinfo {volume} {79}},\
  \bibinfo {pages} {056001} (\bibinfo {year} {2016})}\BibitemShut {NoStop}%
\bibitem [{\citenamefont {Karrasch}\ \emph {et~al.}(2013)\citenamefont
  {Karrasch}, \citenamefont {Ilan},\ and\ \citenamefont
  {Moore}}]{karrasch2013}%
  \BibitemOpen
  \bibfield  {author} {\bibinfo {author} {\bibfnamefont {C.}~\bibnamefont
  {Karrasch}}, \bibinfo {author} {\bibfnamefont {R.}~\bibnamefont {Ilan}},\
  and\ \bibinfo {author} {\bibfnamefont {J.~E.}\ \bibnamefont {Moore}},\
  }\bibfield  {title} {\bibinfo {title} {{\color{black}Nonequilibrium Thermal
  Transport and Its Relation to Linear Response}},\ }\href
  {https://doi.org/10.1103/PhysRevB.88.195129} {\bibfield  {journal} {\bibinfo
  {journal} {Physical Review B}\ }\textbf {\bibinfo {volume} {88}},\ \bibinfo
  {pages} {195129} (\bibinfo {year} {2013})}\BibitemShut {NoStop}%
\bibitem [{\citenamefont {Bernard}\ and\ \citenamefont
  {Doyon}(2012)}]{bernard2012}%
  \BibitemOpen
  \bibfield  {author} {\bibinfo {author} {\bibfnamefont {D.}~\bibnamefont
  {Bernard}}\ and\ \bibinfo {author} {\bibfnamefont {B.}~\bibnamefont
  {Doyon}},\ }\bibfield  {title} {\bibinfo {title} {{\color{black}Energy Flow in
  Non-Equilibrium Conformal Field Theory}},\ }\href
  {https://doi.org/10.1088/1751-8113/45/36/362001} {\bibfield  {journal}
  {\bibinfo  {journal} {Journal of Physics A: Mathematical and Theoretical}\
  }\textbf {\bibinfo {volume} {45}},\ \bibinfo {pages} {362001} (\bibinfo
  {year} {2012})}\BibitemShut {NoStop}%
\bibitem [{\citenamefont {Bhaseen}\ \emph {et~al.}(2015)\citenamefont
  {Bhaseen}, \citenamefont {Doyon}, \citenamefont {Lucas},\ and\ \citenamefont
  {Schalm}}]{bhaseen2015}%
  \BibitemOpen
  \bibfield  {author} {\bibinfo {author} {\bibfnamefont {M.~J.}\ \bibnamefont
  {Bhaseen}}, \bibinfo {author} {\bibfnamefont {B.}~\bibnamefont {Doyon}},
  \bibinfo {author} {\bibfnamefont {A.}~\bibnamefont {Lucas}},\ and\ \bibinfo
  {author} {\bibfnamefont {K.}~\bibnamefont {Schalm}},\ }\bibfield  {title}
  {\bibinfo {title} {{\color{black}Energy Flow in Quantum Critical Systems Far
  from Equilibrium}},\ }\href {https://doi.org/10.1038/nphys3320} {\bibfield
  {journal} {\bibinfo  {journal} {Nature Physics}\ }\textbf {\bibinfo {volume}
  {11}},\ \bibinfo {pages} {509} (\bibinfo {year} {2015})}\BibitemShut
  {NoStop}%
\bibitem [{\citenamefont {Lucas}\ \emph {et~al.}(2016)\citenamefont {Lucas},
  \citenamefont {Schalm}, \citenamefont {Doyon},\ and\ \citenamefont
  {Bhaseen}}]{lucas2016}%
  \BibitemOpen
  \bibfield  {author} {\bibinfo {author} {\bibfnamefont {A.}~\bibnamefont
  {Lucas}}, \bibinfo {author} {\bibfnamefont {K.}~\bibnamefont {Schalm}},
  \bibinfo {author} {\bibfnamefont {B.}~\bibnamefont {Doyon}},\ and\ \bibinfo
  {author} {\bibfnamefont {M.~J.}\ \bibnamefont {Bhaseen}},\ }\bibfield
  {title} {\bibinfo {title} {{\color{black}Shock Waves, Rarefaction Waves, and
  Nonequilibrium Steady States in Quantum Critical Systems}},\ }\href
  {https://doi.org/10.1103/PhysRevD.94.025004} {\bibfield  {journal} {\bibinfo
  {journal} {Physical Review D}\ }\textbf {\bibinfo {volume} {94}},\ \bibinfo
  {pages} {025004} (\bibinfo {year} {2016})}\BibitemShut {NoStop}%
\bibitem [{\citenamefont {Chang}\ \emph {et~al.}(2014)\citenamefont {Chang},
  \citenamefont {Karch},\ and\ \citenamefont {Yarom}}]{chang2014}%
  \BibitemOpen
  \bibfield  {author} {\bibinfo {author} {\bibfnamefont {H.-C.}\ \bibnamefont
  {Chang}}, \bibinfo {author} {\bibfnamefont {A.}~\bibnamefont {Karch}},\ and\
  \bibinfo {author} {\bibfnamefont {A.}~\bibnamefont {Yarom}},\ }\bibfield
  {title} {\bibinfo {title} {{\color{black}An Ansatz for One Dimensional Steady
  State Configurations}},\ }\href
  {https://doi.org/10.1088/1742-5468/2014/06/P06018} {\bibfield  {journal}
  {\bibinfo  {journal} {J. Stat. Mech.}\ }\textbf {\bibinfo {volume} {1406}},\
  \bibinfo {pages} {P06018} (\bibinfo {year} {2014})}\BibitemShut {NoStop}%
\bibitem [{\citenamefont {Amado}\ and\ \citenamefont
  {Yarom}(2015)}]{amado2015}%
  \BibitemOpen
  \bibfield  {author} {\bibinfo {author} {\bibfnamefont {I.}~\bibnamefont
  {Amado}}\ and\ \bibinfo {author} {\bibfnamefont {A.}~\bibnamefont {Yarom}},\
  }\bibfield  {title} {\bibinfo {title} {{\color{black}Black Brane Steady
  States}},\ }\href {https://doi.org/10.1007/JHEP10(2015)015} {\bibfield
  {journal} {\bibinfo  {journal} {JHEP}\ }\textbf {\bibinfo {volume} {10}},\
  \bibinfo {pages} {015 (2015)}}\BibitemShut {NoStop}%
\bibitem [{\citenamefont {Ecker}\ \emph {et~al.}(2021)\citenamefont {Ecker},
  \citenamefont {Erdmenger},\ and\ \citenamefont {van~der Schee}}]{ecker2021}%
  \BibitemOpen
  \bibfield  {author} {\bibinfo {author} {\bibfnamefont {C.}~\bibnamefont
  {Ecker}}, \bibinfo {author} {\bibfnamefont {J.}~\bibnamefont {Erdmenger}},\
  and\ \bibinfo {author} {\bibfnamefont {W.}~\bibnamefont {van~der Schee}},\
  }\bibfield  {title} {\bibinfo {title} {{\color{black}Non-Equilibrium Steady
  State Formation in 3+1 Dimensions}},\ }\href
  {https://doi.org/10.21468/SciPostPhys.11.3.047} {\bibfield  {journal}
  {\bibinfo  {journal} {SciPost Physics}\ }\textbf {\bibinfo {volume} {11}},\
  \bibinfo {pages} {047} (\bibinfo {year} {2021})},\ \Eprint
  {https://arxiv.org/abs/2103.10435} {arXiv:2103.10435 [hep-th]} \BibitemShut
  {NoStop}%
\bibitem [{\citenamefont {Chesler}\ and\ \citenamefont
  {Yaffe}(2014)}]{chesler2014}%
  \BibitemOpen
  \bibfield  {author} {\bibinfo {author} {\bibfnamefont {P.~M.}\ \bibnamefont
  {Chesler}}\ and\ \bibinfo {author} {\bibfnamefont {L.~G.}\ \bibnamefont
  {Yaffe}},\ }\bibfield  {title} {\bibinfo {title} {{\color{black}Numerical
  Solution of Gravitational Dynamics in Asymptotically Anti-de {{Sitter}}
  Spacetimes}},\ }\href {https://doi.org/10.1007/JHEP07(2014)086} {\bibfield
  {journal} {\bibinfo  {journal} {JHEP}\ }\textbf {\bibinfo {volume} {7}},\
  \bibinfo {pages} {86}}\BibitemShut {NoStop}%
\bibitem [{\citenamefont {Hubeny}\ and\ \citenamefont
  {Rangamani}(2010)}]{hubeny2010}%
  \BibitemOpen
  \bibfield  {author} {\bibinfo {author} {\bibfnamefont {V.~E.}\ \bibnamefont
  {Hubeny}}\ and\ \bibinfo {author} {\bibfnamefont {M.}~\bibnamefont
  {Rangamani}},\ }\bibfield  {title} {\bibinfo {title} {{\color{black}A
  Holographic View on Physics out of Equilibrium}},\ }\href
  {https://doi.org/10.1155/2010/297916} {\bibfield  {journal} {\bibinfo
  {journal} {Advances in High Energy Physics}\ }\textbf {\bibinfo {volume}
  {2010}},\ \bibinfo {pages} {1} (\bibinfo {year} {2010})},\ \Eprint
  {https://arxiv.org/abs/1006.3675} {arXiv:1006.3675 [gr-qc]} \BibitemShut
  {NoStop}%
\bibitem [{\citenamefont {Liu}\ and\ \citenamefont {Sonner}(2018)}]{liu2018}%
  \BibitemOpen
  \bibfield  {author} {\bibinfo {author} {\bibfnamefont {H.}~\bibnamefont
  {Liu}}\ and\ \bibinfo {author} {\bibfnamefont {J.}~\bibnamefont {Sonner}},\
  }\href@noop {} {\bibinfo {title} {{\color{black}Holographic Systems Far from
  Equilibrium: A Review}}} (\bibinfo {year} {2018}),\ \Eprint
  {https://arxiv.org/abs/1810.02367} {arXiv:1810.02367 [hep-th]} \BibitemShut
  {NoStop}%
\bibitem [{\citenamefont {Cardoso}\ \emph {et~al.}(2012)\citenamefont
  {Cardoso}, \citenamefont {Gualtieri}, \citenamefont {Herdeiro}, \citenamefont
  {Sperhake} \emph {et~al.}}]{cardoso2012}%
  \BibitemOpen
  \bibfield  {author} {\bibinfo {author} {\bibfnamefont {V.}~\bibnamefont
  {Cardoso}}, \bibinfo {author} {\bibfnamefont {L.}~\bibnamefont {Gualtieri}},
  \bibinfo {author} {\bibfnamefont {C.}~\bibnamefont {Herdeiro}}, \bibinfo
  {author} {\bibfnamefont {U.}~\bibnamefont {Sperhake}}, \emph {et~al.},\
  }\bibfield  {title} {\bibinfo {title} {{\color{black}{{NR}}/{{HEP}}: Roadmap
  for the Future}},\ }\href {https://doi.org/10.1088/0264-9381/29/24/244001}
  {\bibfield  {journal} {\bibinfo  {journal} {Classical and Quantum Gravity}\
  }\textbf {\bibinfo {volume} {29}},\ \bibinfo {pages} {244001} (\bibinfo
  {year} {2012})},\ \Eprint {https://arxiv.org/abs/1201.5118} {arXiv:1201.5118
  [astro-ph]} \BibitemShut {NoStop}%
\bibitem [{\citenamefont {Fujita}\ \emph {et~al.}(2011)\citenamefont {Fujita},
  \citenamefont {Takayanagi},\ and\ \citenamefont {Tonni}}]{fujita2011}%
  \BibitemOpen
  \bibfield  {author} {\bibinfo {author} {\bibfnamefont {M.}~\bibnamefont
  {Fujita}}, \bibinfo {author} {\bibfnamefont {T.}~\bibnamefont {Takayanagi}},\
  and\ \bibinfo {author} {\bibfnamefont {E.}~\bibnamefont {Tonni}},\ }\bibfield
   {title} {\bibinfo {title} {{\color{black}Aspects of {{AdS}}/{{BCFT}}}},\
  }\href {https://doi.org/10.1007/JHEP11(2011)043} {\bibfield  {journal}
  {\bibinfo  {journal} {JHEP}\ }\textbf {\bibinfo {volume} {11}},\ \bibinfo
  {pages} {043 (2011)}}\BibitemShut {NoStop}%
\bibitem [{\citenamefont {Nozaki}\ \emph {et~al.}(2012)\citenamefont {Nozaki},
  \citenamefont {Takayanagi},\ and\ \citenamefont {Ugajin}}]{nozaki2012}%
  \BibitemOpen
  \bibfield  {author} {\bibinfo {author} {\bibfnamefont {M.}~\bibnamefont
  {Nozaki}}, \bibinfo {author} {\bibfnamefont {T.}~\bibnamefont {Takayanagi}},\
  and\ \bibinfo {author} {\bibfnamefont {T.}~\bibnamefont {Ugajin}},\
  }\bibfield  {title} {\bibinfo {title} {{\color{black}Central {{Charges}} for
  {{BCFTs}} and {{Holography}}}},\ }\href
  {https://doi.org/10.1007/JHEP06(2012)066} {\bibfield  {journal} {\bibinfo
  {journal} {JHEP}\ }\textbf {\bibinfo {volume} {06}},\ \bibinfo {pages} {066
  (2012)}}\BibitemShut {NoStop}%
\bibitem [{\citenamefont {Takayanagi}(2011)}]{takayanagi2011}%
  \BibitemOpen
  \bibfield  {author} {\bibinfo {author} {\bibfnamefont {T.}~\bibnamefont
  {Takayanagi}},\ }\bibfield  {title} {\bibinfo {title}
  {{\color{black}Holographic {{Dual}} of {{BCFT}}}},\ }\href
  {https://doi.org/10.1103/PhysRevLett.107.101602} {\bibfield  {journal}
  {\bibinfo  {journal} {Phys. Rev. Lett.}\ }\textbf {\bibinfo {volume} {107}},\
  \bibinfo {pages} {101602} (\bibinfo {year} {2011})}\BibitemShut {NoStop}%
\bibitem [{\citenamefont {Andrei}\ \emph {et~al.}(2020)\citenamefont {Andrei}
  \emph {et~al.}}]{Andrei:2018die}%
  \BibitemOpen
  \bibfield  {author} {\bibinfo {author} {\bibfnamefont {N.}~\bibnamefont
  {Andrei}} \emph {et~al.},\ }\bibfield  {title} {\bibinfo {title}
  {{\color{black}{Boundary and Defect CFT: Open Problems and Applications}}},\
  }\href {https://doi.org/10.1088/1751-8121/abb0fe} {\bibfield  {journal}
  {\bibinfo  {journal} {J. Phys. A}\ }\textbf {\bibinfo {volume} {53}},\
  \bibinfo {pages} {453002} (\bibinfo {year} {2020})},\ \Eprint
  {https://arxiv.org/abs/1810.05697} {arXiv:1810.05697 [hep-th]} \BibitemShut
  {NoStop}%
\bibitem [{\citenamefont {Izumi}\ \emph {et~al.}(2022)\citenamefont {Izumi},
  \citenamefont {Shiromizu}, \citenamefont {Suzuki}, \citenamefont
  {Takayanagi},\ and\ \citenamefont {Tanahashi}}]{izumi2022}%
  \BibitemOpen
  \bibfield  {author} {\bibinfo {author} {\bibfnamefont {K.}~\bibnamefont
  {Izumi}}, \bibinfo {author} {\bibfnamefont {T.}~\bibnamefont {Shiromizu}},
  \bibinfo {author} {\bibfnamefont {K.}~\bibnamefont {Suzuki}}, \bibinfo
  {author} {\bibfnamefont {T.}~\bibnamefont {Takayanagi}},\ and\ \bibinfo
  {author} {\bibfnamefont {N.}~\bibnamefont {Tanahashi}},\ }\bibfield  {title}
  {\bibinfo {title} {{\color{black}Brane Dynamics of Holographic {{BCFTs}}}},\
  }\href {https://doi.org/10.1007/JHEP10(2022)050} {\bibfield  {journal}
  {\bibinfo  {journal} {JHEP}\ }\textbf {\bibinfo {volume} {10}},\ \bibinfo
  {pages} {050 (2022)}}\BibitemShut {NoStop}%
\bibitem [{\citenamefont {Engelhardt}\ and\ \citenamefont
  {Wall}(2018)}]{engelhardt2018}%
  \BibitemOpen
  \bibfield  {author} {\bibinfo {author} {\bibfnamefont {N.}~\bibnamefont
  {Engelhardt}}\ and\ \bibinfo {author} {\bibfnamefont {A.~C.}\ \bibnamefont
  {Wall}},\ }\bibfield  {title} {\bibinfo {title} {{\color{black}Decoding the
  {{Apparent Horizon}}: {{Coarse-Grained Holographic Entropy}}}},\ }\href
  {https://doi.org/10.1103/PhysRevLett.121.211301} {\bibfield  {journal}
  {\bibinfo  {journal} {Physical Review Letters}\ }\textbf {\bibinfo {volume}
  {121}},\ \bibinfo {pages} {211301} (\bibinfo {year} {2018})}\BibitemShut
  {NoStop}%
\bibitem [{\citenamefont {Baggioli}\ \emph {et~al.}(2022)\citenamefont
  {Baggioli}, \citenamefont {Li},\ and\ \citenamefont
  {Sun}}]{Baggioli:2021tzr}%
  \BibitemOpen
  \bibfield  {author} {\bibinfo {author} {\bibfnamefont {M.}~\bibnamefont
  {Baggioli}}, \bibinfo {author} {\bibfnamefont {L.}~\bibnamefont {Li}},\ and\
  \bibinfo {author} {\bibfnamefont {H.-T.}\ \bibnamefont {Sun}},\ }\bibfield
  {title} {\bibinfo {title} {{\color{black}{Shear Flows in Far-from-Equilibrium
  Strongly Coupled Fluids}}},\ }\href
  {https://doi.org/10.1103/PhysRevLett.129.011602} {\bibfield  {journal}
  {\bibinfo  {journal} {Phys. Rev. Lett.}\ }\textbf {\bibinfo {volume} {129}},\
  \bibinfo {pages} {011602} (\bibinfo {year} {2022})},\ \Eprint
  {https://arxiv.org/abs/2112.14855} {arXiv:2112.14855 [hep-th]} \BibitemShut
  {NoStop}%
\bibitem [{\citenamefont {Hollands}\ \emph {et~al.}(2024)\citenamefont
  {Hollands}, \citenamefont {Wald},\ and\ \citenamefont
  {Zhang}}]{hollands2024}%
  \BibitemOpen
  \bibfield  {author} {\bibinfo {author} {\bibfnamefont {S.}~\bibnamefont
  {Hollands}}, \bibinfo {author} {\bibfnamefont {R.~M.}\ \bibnamefont {Wald}},\
  and\ \bibinfo {author} {\bibfnamefont {V.~G.}\ \bibnamefont {Zhang}},\
  }\bibfield  {title} {\bibinfo {title} {{\color{black}Entropy of Dynamical
  Black Holes}},\ }\href {https://doi.org/10.1103/PhysRevD.110.024070}
  {\bibfield  {journal} {\bibinfo  {journal} {Physical Review D}\ }\textbf
  {\bibinfo {volume} {110}},\ \bibinfo {pages} {024070} (\bibinfo {year}
  {2024})}\BibitemShut {NoStop}%
\bibitem [{\citenamefont {Rougemont}\ \emph {et~al.}(2022)\citenamefont
  {Rougemont}, \citenamefont {Barreto},\ and\ \citenamefont
  {Noronha}}]{rougemont2022}%
  \BibitemOpen
  \bibfield  {author} {\bibinfo {author} {\bibfnamefont {R.}~\bibnamefont
  {Rougemont}}, \bibinfo {author} {\bibfnamefont {W.}~\bibnamefont {Barreto}},\
  and\ \bibinfo {author} {\bibfnamefont {J.}~\bibnamefont {Noronha}},\
  }\bibfield  {title} {\bibinfo {title} {{\color{black}Hydrodynamization Times
  of a Holographic Fluid Far from Equilibrium}},\ }\href
  {https://doi.org/10.1103/PhysRevD.105.046009} {\bibfield  {journal} {\bibinfo
   {journal} {Physical Review D}\ }\textbf {\bibinfo {volume} {105}},\ \bibinfo
  {pages} {046009} (\bibinfo {year} {2022})}\BibitemShut {NoStop}%
\bibitem [{\citenamefont {Baibhav}\ \emph {et~al.}(2023)\citenamefont
  {Baibhav}, \citenamefont {Cheung}, \citenamefont {Berti}, \citenamefont
  {Cardoso}, \citenamefont {Carullo}, \citenamefont {Cotesta}, \citenamefont
  {Del~Pozzo},\ and\ \citenamefont {Duque}}]{baibhav2023}%
  \BibitemOpen
  \bibfield  {author} {\bibinfo {author} {\bibfnamefont {V.}~\bibnamefont
  {Baibhav}}, \bibinfo {author} {\bibfnamefont {M.~H.-Y.}\ \bibnamefont
  {Cheung}}, \bibinfo {author} {\bibfnamefont {E.}~\bibnamefont {Berti}},
  \bibinfo {author} {\bibfnamefont {V.}~\bibnamefont {Cardoso}}, \bibinfo
  {author} {\bibfnamefont {G.}~\bibnamefont {Carullo}}, \bibinfo {author}
  {\bibfnamefont {R.}~\bibnamefont {Cotesta}}, \bibinfo {author} {\bibfnamefont
  {W.}~\bibnamefont {Del~Pozzo}},\ and\ \bibinfo {author} {\bibfnamefont
  {F.}~\bibnamefont {Duque}},\ }\bibfield  {title} {\bibinfo {title}
  {{\color{black}Agnostic Black Hole Spectroscopy: {{Quasinormal}} Mode Content
  of Numerical Relativity Waveforms and Limits of Validity of Linear
  Perturbation Theory}},\ }\href {https://doi.org/10.1103/PhysRevD.108.104020}
  {\bibfield  {journal} {\bibinfo  {journal} {Physical Review D}\ }\textbf
  {\bibinfo {volume} {108}},\ \bibinfo {pages} {104020} (\bibinfo {year}
  {2023})}\BibitemShut {NoStop}%
\bibitem [{\citenamefont {Cheung}\ \emph {et~al.}(2023)\citenamefont {Cheung}
  \emph {et~al.}}]{cheung2023}%
  \BibitemOpen
  \bibfield  {author} {\bibinfo {author} {\bibfnamefont {M.~H.-Y.}\
  \bibnamefont {Cheung}} \emph {et~al.},\ }\bibfield  {title} {\bibinfo {title}
  {{\color{black}Nonlinear {{Effects}} in {{Black Hole Ringdown}}}},\ }\href
  {https://doi.org/10.1103/PhysRevLett.130.081401} {\bibfield  {journal}
  {\bibinfo  {journal} {Phys. Rev. Lett.}\ }\textbf {\bibinfo {volume} {130}},\
  \bibinfo {pages} {081401} (\bibinfo {year} {2023})}\BibitemShut {NoStop}%
\bibitem [{\citenamefont {Cheung}\ \emph {et~al.}(2024)\citenamefont {Cheung},
  \citenamefont {Berti}, \citenamefont {Baibhav},\ and\ \citenamefont
  {Cotesta}}]{cheung2024}%
  \BibitemOpen
  \bibfield  {author} {\bibinfo {author} {\bibfnamefont {M.~H.-Y.}\
  \bibnamefont {Cheung}}, \bibinfo {author} {\bibfnamefont {E.}~\bibnamefont
  {Berti}}, \bibinfo {author} {\bibfnamefont {V.}~\bibnamefont {Baibhav}},\
  and\ \bibinfo {author} {\bibfnamefont {R.}~\bibnamefont {Cotesta}},\
  }\bibfield  {title} {\bibinfo {title} {{\color{black}Extracting Linear and
  Nonlinear Quasinormal Modes from Black Hole Merger Simulations}},\ }\href
  {https://doi.org/10.1103/PhysRevD.109.044069} {\bibfield  {journal} {\bibinfo
   {journal} {Phys. Rev. D}\ }\textbf {\bibinfo {volume} {109}},\ \bibinfo
  {pages} {044069} (\bibinfo {year} {2024})}\BibitemShut {NoStop}%
\end{thebibliography}%
\newpage
\onecolumngrid
\appendix 
\clearpage
\renewcommand\thefigure{S\arabic{figure}}    
\setcounter{figure}{0} 
\renewcommand{\theequation}{S\arabic{equation}}
\setcounter{equation}{0}
\renewcommand{\thesubsection}{SI\arabic{subsection}}
\section*{Supplementary Information}
\subsection{The holographic  setup}\label{holographic setup}
We consider a four-dimensional holographic setup described by the action: 
\begin{equation}\label{action}
S\,=\frac{1}{2\kappa_N^2}\,\int d^4x \sqrt{-g}
\left[\mathcal{R}-2 \Lambda\right]\,,
\end{equation}
where $\kappa_N^2=8\pi G_N$ is the 4D gravitational Newton’s constant,  $\Lambda=-3$ is the negative cosmological constant.

To capture the non-trivial dynamics of thermalization between  two systems with different temperatures at the boundary, we consider the following metric in Eddington-Finkelstein (EF) coordinates:
\begin{equation}\label{appansatz}
ds^2=-2A(u,t,x)dt^2-\frac{2 \,du\, dt}{u^2}-2F(u,t,x)dtdx+\Sigma(u,t,x)^2  (e^{B(u,t,x)}dx^2+e^{-B(u,t,x)}dy^2)\,,
\end{equation}
where the metric depends on the holographic radial coordinate $u$, time coordinate $t$ and one spatial coordinate $x$.
In our notations, the asymptotic AdS boundary is located at $u=0$. 
Rather than using the event horizon -- which depends on the complete spacetime history -- we adopt the apparent horizon at $u=u_A$ as our interior boundary in non-equilibrium scenarios. This corresponds to the outermost trapped null surface that forms behind the event horizon.

We use Chesler and Yaffe's formalism to solve the problem \cite{chesler2014}. The bulk equations of motion for our system are given by 
\begin{align}
u\Sigma{}^{''} +2\Sigma{}^{'}+\frac{{u B{}^{'}}^2 }{4}&\Sigma=0\,,\label{eqS}\\
F''+\frac{2- B 'u}{u}F'+&\left( \frac{B '^2}{2}- B ''-\frac{2 \Sigma '^2}{ \Sigma ^2}-\frac{2 B '( \Sigma + \Sigma 'u)}{ \Sigma u}\right)F=-\frac{-2\partial_x \Sigma '+2 B '\partial_x \Sigma + \Sigma \partial_x B' - B '\partial_x B  \Sigma }{ \Sigma u^2}-\frac{2 \Sigma '\partial_x \Sigma }{ \Sigma ^2u^2}\,, \label{eqF}\\
{d_+\Sigma }'+\frac{ \Sigma'{d_+\Sigma }}{\Sigma }=
&\frac{e^{-B}}{8 \Sigma ^3 u^2} \Big(-2 F \Sigma ^2 u^2 \left(u^2 B' F'-B_x'\right)+2 B_x \Sigma ^2 u^2 \left(F'-F B'\right)+2 F_x \Sigma ^2 u^2 B'-F^2 \Sigma ^2 u^4 \left(B'\right)^2-12 e^B \Sigma ^4\notag\\
&-2 B_{xx} \Sigma ^2-4 \Sigma  \left(-\Sigma _{xx}+u^2 \Sigma ' \left(-B_x F+F_x+2 F^2 u-F u^2 F'\right)+B_x \Sigma _x+F u^2 \Sigma _x'+F^2 u^4 \Sigma ''\right)\notag\\
&+2 B_x^2 \Sigma ^2-2 \Sigma ^2 u^2 F_x'-4 \Sigma _x \left(\Sigma _x-F u^2 \Sigma '\right)+\Sigma ^2 u^4 \left(F'\right)^2\Big)
\,, 
\label{eqdpS}\\
d_+B'+\frac{d_+B \Sigma '}{\Sigma }=&-\frac{e^{-B}}{4 \Sigma ^4} \Big{(}4 e^B d_+\Sigma  \Sigma ^3 B'+F \left(-2 \Sigma  B' \left(B_x \Sigma -2 \Sigma _x+\Sigma  u^2 F'\right)+2 \Sigma ^2 B_x'-4 \Sigma  \Sigma _x'+4 \Sigma _x \Sigma '+4 \Sigma  u^2 F' \Sigma '\right)\notag\\
&+2 F^2 u \left(\Sigma  u \left(2 \Sigma ''-\Sigma  B''\right)+\Sigma ^2 u \left(B'\right)^2-2 \Sigma  B' \left(\Sigma +u \Sigma '\right)+4 \Sigma  \Sigma '-2 u \left(\Sigma '\right)^2\right)+\Sigma ^2 \left(2 F_x'-u^2 \left(F'\right)^2\right)\notag\\
&-4 \Sigma _x \Sigma  F'\Big{)}
\,, 
\label{eqdpB}\\
\frac{u A''}{2}+A'=&-\frac{e^{-B}}{4 \Sigma ^4 u^3} \Big(-e^B d_+B \Sigma ^4 u^2 B'+2 F \Sigma  u^2 \left(-\Sigma _x B'+F u \left(\Sigma ' \left(u B'-2\right)-u \Sigma ''\right)+\Sigma _x'\right)\notag\\
&+\Sigma ^2 u^2 \left(F \left(B_x B'-B_x'\right)+F^2 u \left(u B''-u \left(B'\right)^2+2 B'\right)+B_x F'-F_x'+u^2 \left(F'\right)^2\right)\notag\\
&-4 e^B \Sigma ^3 u^2 d_+\Sigma '-6 e^B \Sigma ^4+2 F u^2 \Sigma ' \left(F u^2 \Sigma '-\Sigma _x\right)\Big)
\,.
\label{eqA}
\end{align}
There are five more equations of motion from the Einstein equations, including one second-order time derivative equation (see~\eqref{eq:cons2order} below) and four first-order time derivative equations (constraint equations). We will check that all of them are satisfied in our numerics.

In the expressions above, we have introduced the directional derivative 
$d_+\mathcal{F}:= \dot{\mathcal{F}}-\frac{u^2 A}{2}\mathcal{F}^{'}$, where the prime ($'$)  denotes differentiation with respect to the radial coordinate $u$,  and the dot ($\dot{}$) denotes the time derivative $\partial_t$. Additionally,  $\mathcal{F}_x\equiv \partial_x \mathcal{F}$ denotes the spatial derivative with respect to $x$,  $\mathcal{F}_{xx}\equiv \partial_x^2 \mathcal{F}$ is the second-order spatial derivative with  respect to $x$.

 The boundary asymptotics for the different bulk fields are found to be
\begin{equation}\label{bdyexpansion}
\begin{split}
&A=\frac{1}{2u^2}\left[1+2s_1u+\left( s_1^2-2\dot{s}_1\right)u^2+a_3 u^3+\mathcal{O}(u^4)\right]\,,\\
&\Sigma=\frac{1}{u}\left[1+s_1 u+\mathcal{O}(u^3)\right]\,,\\
&F=-\partial_x s_1+f_3u-(4f_3s_1+3\partial_xb_3)u^2+\mathcal{O}(u^3)\,,\\
&B=b_3 u^3+\mathcal{O}(u^4)\,.
\end{split}
\end{equation}
Without loss of generality, we fix the apparent horizon at $u_A=1$ using the  residual diffeomorphism symmetry and we keep $s_1$ as a dynamical parameter.\footnote{Note that the EF coordinates possess  residual diffeomorphism invariance under the transformation $u\rightarrow \tilde{u}=\frac{u}{1+\lambda(t)u}$  with $\lambda(t)$ an arbitrary function of time. This transformation induces the shift $s_1\rightarrow\tilde{s}_1=s_1-\lambda(t)$.}
By choosing the residual diffeomorphism parameter $s_1$, we can fix the apparent horizon located at $u_h=1$. 
To preserve this fixed horizon position during time evolution, we must 
solve the following  constraint   equation for $A$ at the apparent horizon.
\begin{align}
    &A_{xx}+\alpha_a A_x+\beta_a A+\gamma_a=0\,,\label{eqAh}
\end{align}
where these parameters are given by 
\begin{align}
        \alpha_a=&\left(-\left(u^2 \left(F B'-F'\right)\right)-B_x-\frac{2 F u^2 \Sigma '}{\Sigma }\right)\,,\\
    \beta_a=&\frac{1}{4 \Sigma ^2} \Big(\Sigma ^2 \left(u^2 \left(-2 F_x B'+u^2 \left(F'-F B'\right)^2-2 F B_x'+2 F_x'\right)+2 B_x u^2 \left(F B'-F'\right)-2 B_{xx}+2 B_x^2\right)\nonumber\\
    &-4 \Sigma  \left(u^2 \left(\Sigma ' \left(F u^2 \left(F'-F B'\right)+F_x\right)+F \Sigma _x'\right)-\Sigma _{xx}+B_x \left(\Sigma _x-F u^2 \Sigma '\right)\right)\nonumber\\
    &-12 e^B \Sigma ^4+4 \left(-\Sigma _x^2+\Sigma _x F u^2 \Sigma '+F^2 u^4 \left(\Sigma '\right)^2\right)\Big)\,,\\
    \gamma_a=&\frac{e^{-B} F}{2 \Sigma ^2} \left(F \left(u^2 \left(F_x'-2 F_x B'\right)-B_{xx}+B_x^2-2 B_x u^2 F'\right)+F^2 u^2 \left(2 B_x B'-B_x'\right)+2 \left(F_{xx}+F_x \left(u^2 F'-2 B_x\right)\right)\right)\nonumber\\
    &+\frac{e^{-B} F}{2 \Sigma ^3} \left(F u^2 \Sigma ' \left(F \left(u^2 \left(F B'-2 F'\right)+2 B_x\right)-4 F_x\right)-2 \left(F \left(\Sigma _{xx}+\Sigma _x \left(u^2 F'-3 B_x\right)\right)+2 F_x \Sigma _x\right)\right)\nonumber\\
    &+\frac{1}{2} d_+B \left(F \left(F u^2 B'+\frac{4 \Sigma _x}{\Sigma }\right)-2 F_x\right)-\frac{1}{2} e^B d_+B^2 \Sigma ^2-2 e^B d_+\Sigma ^2+\frac{e^{-B} F^2 \left(\Sigma _x+F u^2 \Sigma '\right)^2}{\Sigma ^4}+\frac{F^2 u^2 d_+\Sigma '}{\Sigma }\nonumber\\
    &+\frac{d_+\Sigma  }{\Sigma ^2}\left(2 \Sigma  (B_x F-F_x)+F^2 u^2 \Sigma '\right)\,.
\end{align}

The renormalized action is obtained by adding the standard Gibbons-Hawking term and appropriate counterterms: 
\begin{equation}
S_{\text{ren}}=S+\frac{1}{2\kappa_N^2}\int_\partial\sqrt{-h}\left(2\mathcal{K}+4-\mathcal{R}^h\right)\,,
\end{equation}
where $h_{ab}$ is the induced metric on the boundary, $\mathcal{R}^h$ is the associated Ricci scalar and $\mathcal{K}$ is the trace of the extrinsic curvature which is given by  $\mathcal{K}_{\mu\nu}={h_\mu}^\lambda {h_\nu}^\sigma\nabla_\lambda n_\sigma$ with $n^\mu$ the outward pointing unit normal to the boundary. The boundary stress tensor is then given by
\begin{equation}\label{Htensor}
T_{ab}=\lim_{u\rightarrow 0}-\frac{2}{u\sqrt{-h}}\frac{\delta S_{\text{ren}}}{\delta h^{ab}}=\frac{1}{2\kappa_N^2}\lim_{u\rightarrow 0}\frac{2}{u}\,\left(\mathcal{K} \,h_{ab}-\mathcal{K}_{ab}-2h_{ab}+G^h_{ab}\right)\,,
\end{equation}
with $G^h_{ab}$ the Einstein tensor in terms of the boundary metric $h_{ab}.$
Here, Latin indices run through boundary directions,  \emph{i.e.} $a=t, x, y$; Greek indices run over the bulk spacetime dimensions, \emph{i.e.} $\mu=u,t, x, y$.  

The non-vanishing components of $T_{ab}$ and $J^a$ 
in the dual field theory are 
\begin{eqnarray}
&&{E}\equiv T_{tt} ={ -2a_3}\,,\label{stress0}\\
&&J\equiv T_{tx}=-\frac{3}{2}f_3\,,\\
&&T_{xx} =-a_3+\frac{3}{2}b_3\,,\\
&&T_{yy} =-a_3-\frac{3}{2}b_3\,,
\end{eqnarray}
where ${E}$ and $J$ are the energy density and momentum density of the dual field theory, respectively. Here, we have chosen $\kappa_N^2=1$.
The boundary metric, in which the dual field theory lives, reads
\begin{equation}\label{Fmetric}
\begin{split}
d s^2= \eta_{ab}dx^a dx^b
=-d t^2+dx^2+dy^2\,.
\end{split}
\end{equation}
 As a double check, one can easily show that the following Ward identities hold.
\begin{equation}\label{wald}
 \eta_{ab} T^{ab} =0,\quad \partial_a {T^{a}}_b=0\,.
\end{equation}
Moreover, the entropy density $\mathcal{S}=2\pi \Sigma(u_A,t)^2$ is   associated with the apparent horizon at $u=u_A$. While in equilibrium black hole thermodynamics the entropy density is defined by the event horizon's area, for non-equilibrium scenarios a more appropriate one is the apparent horizon, see \emph{e.g}~\cite{engelhardt2018,Baggioli:2021tzr, hollands2024, rougemont2022}. 


Within the AdS/BCFT framework, a finite-size field theory can be holographically described by a 
sub-region of asymptotically AdS spacetime.  
The embedding of this subregion's boundary in AdS spacetime, a bulk brane denoted as $Q$, is determined by imposing specific conditions on the Brown-York tensor evaluated on $Q$. In our setup, the spatial slice of the system is shown in Fig. S1. We study a strip shape boundary system.

\begin{figure}[h]
	 \centering
	  \includegraphics[width=0.5\textwidth]{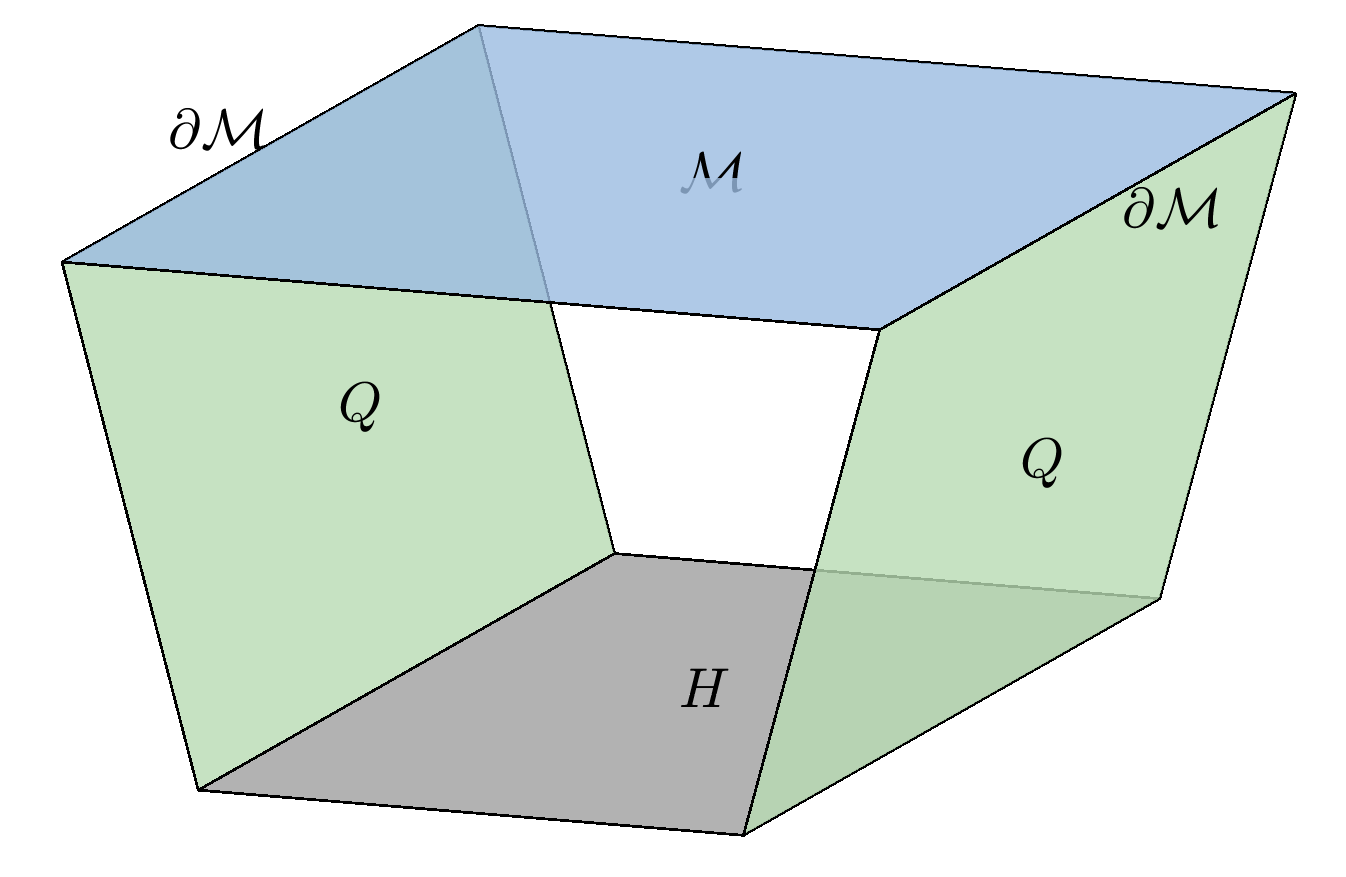}
	    \caption{Illustration for the space slice of the bulk system. $\mathcal{M}$ is the boundary CFT, which is $2$ spatial dimensions. The boundary of $\mathcal{M}$ extends into the bulk AdS spacetime and forms two brane $Q$. $H$ denotes the apparent horizon of the  black hole.}
	    \label{fig:spacesclice}
\end{figure}

We consider a tensionless bulk brane $Q$ carrying no additional matter field. The induced metric on the brane $Q$ is 
\begin{align}
    \gamma_{ab}=g_{ab}-m_am_b\,,
\end{align}
where $m^a=\frac{e^{B/2}F}{\Sigma}\left(\frac{\partial}{\partial u}\right)^a+\frac{e^{B/2}}{\Sigma}\left(\frac{\partial}{\partial x}\right)^a$ is the unit normal vector to the brane. 
With Neumann boundary condition on $Q$, the energy momentum tensor on the brane satisfies 
\begin{align}
\label{eq:hatT=0}
    \hat{T}_{ab}=K_{ab}-K\gamma_{ab}\,,
\end{align}
where $K_{ab}=\gamma_{a}{}^c\nabla_cm_b$ is the extrinsic curvature of bulk brane. The non-trivial components of $\hat{T}_{ab}$ are
\begin{align}
    \hat{T}_{tu}&=\frac{e^{-b/2}}{2\Sigma^2}\left(-F \Sigma  u^2 B'+\partial_xB \Sigma -2 \partial_x \Sigma+\Sigma  u^2 F'+2 F u^2 \Sigma '\right)\,,\label{eq:S'}\\
    \hat{T}_{uu}&=\frac{e^{-b/2}}{\Sigma^2}\left(\Sigma  \left(F u^2 A'-A F u^2 B'+A \partial_xB- \partial_xA+\dot{F}\right)-2 A \left(\partial_x \Sigma-F u^2 \Sigma '\right)\right)\,,\label{eq:braneT11}\\
    \hat{T}_{yy}&=-\frac{F'e^{-B/2}}{\Sigma^3}\,.\label{eq:F'}
\end{align}
The condition of vanishing energy-momentum tensor on the brane yields the gravitational boundary conditions for the bulk metric.

To study a finite-size isolated system, we impose the physical requirement that the energy current $J=0$ at the boundary of field theory. This ensures no energy flow across the boundary. To maintain this condition throughout the time evolution, the time derivative of the energy current must vanish. From the conservation of energy-momentum tensor, we find
\begin{align}
    \partial_aT^{ax}|_{\partial \mathcal{M}}=\partial_t T^{tx}+\partial_x T^{xx}=0
\end{align}
where $T^{yy}$ is homogeneous in the other spatial direction, so the partial derivative vanishes.
The conservation equation imposes a constraint  for $T^{xx}$: 
\begin{align}
    -\partial_x a_3|_{\partial \mathcal{M}}+\frac{3}{2}\partial_x b_3|_{\partial \mathcal{M}}=0\,.\label{eq:consT11}
\end{align}
These boundary conditions at AdS boundary must be extended 
into the bulk. This is achieved by introducing an end-of-world brane, on which we impost the boundary conditions for metric fields. 
Specifically, we require 
$\hat{T}_{ab}=0$ in \eqref{eq:hatT=0} on the  bulk brane. To satisfy this, we adjust the dynamical quantities $B,J$ and the radial gauge $s_1$. The choice of $s_1$ is tied to the condition imposed on $a_3$. 

For the evolution quantities, if we ignore the $\hat{T}_{ab}=0$ condition --- in other words, if we allow for the present of exotic matter field on the brane (which can describe an open system) --- we can choose different (maybe potentially infinitely many) boundary conditions for the bulk brane that are consistent with the initial data and equations of motion (the physical conditions). The choice of  boundary conditions determines the solutions, and among these solutions, the condition  $\hat{T}_{ab}=0$ selects a subclass of systems.

First, consider \eqref{eq:F'}. To ensure this equation vanished on the brane while maintaining smoothness of all functions, we  require
\begin{align}
 F'(u)|_Q=0\,.
\end{align}
From the boundary expansions, this implies 
\begin{align}
    J|_{\partial \mathcal{M}}=0\,,\quad F(u)|_Q=-\partial_x s_1\,,\quad F''(u)|_Q=0\,.
\end{align}
Then we examine the asymptotic solution for $F$, which leads to
\begin{align}
    \partial_xb_3|_{\partial \mathcal{M}}=0\,, \quad \partial_xa_3|_{\partial \mathcal{M}}=0\,,
\end{align}
where the latter follows from \eqref{eq:consT11}. We now see how to choose boundary conditions for $s_1$ and $B$ to satisfy these constraints.

Next, we consider \eqref{eq:S'}. The condition  $\hat{T}^{tu}|_Q=0$ is equivalent to 
\begin{align}
    u^2F(\frac{2\Sigma'}{\Sigma}-B')=\frac{2\partial_x\Sigma}{\Sigma}-\partial_x B\,. \label{eq:Brane11}
\end{align}
Eq.~\eqref{eqS} indicates that the left-hand  side depends only on $B'$ and $\partial_x s_1$, while the right-hand side additionally depends on 
$\partial_x B$. This equation must hold during time evolution and can be regard as a boundary condition for $\partial_x B$. However, another condition for $\partial_x B$ arises from~\eqref{eqF}, 
\begin{align}
   \left( \frac{B '^2}{2}- B ''-\frac{2 \Sigma '^2}{ \Sigma ^2}-\frac{2 B '( \Sigma + \Sigma 'u)}{ \Sigma u}\right)F=-\frac{-2\partial_x \Sigma '+2 B '\partial_x \Sigma + \Sigma \partial_x B' - B '\partial_x B  \Sigma }{ \Sigma u^2}-\frac{2 \Sigma '\partial_x \Sigma }{ \Sigma ^2u^2}\,.\label{eq:Brane11_consist}
\end{align}
These two equations are consistent. To verify this, one may combine  \eqref{eq:Brane11}$'$-$B'$\eqref{eq:Brane11}  with \eqref{eqS}, which yields  \eqref{eq:Brane11_consist}. 

We now examine whether \eqref{eq:braneT11} can be consistently satisfied. This equation is  excepted to provide the boundary condition for $s_1$. However, this is non-trivial: since \eqref{eq:braneT11} must hold on the brane, the adjustable parameters are limited, resulting in insufficient degree of freedom. Thus, this imposes an additional constraint on the initial conditions. Eq.~\eqref{eq:braneT11} is equivalent to 
\begin{align}
    \dot{F}=\partial_x A+F u^2 A'\,.
\end{align}
From the asymptotic behavior near the AdS boundary and the fact that $\dot{F}$ is  constant along the radial coordinate,  we get
\begin{align}
    \dot{F}=-\partial_x \dot{s}_1\,.
\end{align}
Consequently, on the brane, the following must hold during evolution: 
\begin{align}
    \partial_x A+F u^2 A'=\dot{F}\label{eq:addcons}\,,
\end{align}
under the boundary conditions
\begin{align}
    J|_{\partial \mathcal{M}}=0,\quad \partial_x s_1|_{\partial \mathcal{M}}=-F,\quad \partial_x B|_Q =\frac{2\partial_x\Sigma}{\Sigma}-u^2F(\frac{2\Sigma'}{\Sigma}-B')\,,\label{eq:sumbranecond}
\end{align}
where $F$ is only a function of $t$. 

There is no general guarantee that the boundary condition \eqref{eq:sumbranecond} can make the additional constraint of no matter on the tensionless brane \eqref{eq:addcons} satisfied, because \eqref{eq:addcons} depends on the higher derivatives not controlled by \eqref{eq:sumbranecond}. Whether~\eqref{eq:addcons} holds is determined by the initial conditions and equations of motion. 
Although the equations of motion are fixed, 
the uniqueness of the initial conditions remain unclear.  
A fully analysis of the nonlinear equations of motion with infinite degree of freedom is required for a general proof. 

However, one class of initial conditions  satisfies the requirements. This initial conditions give the profile of $J$, $s_1$ and $B$, which determine all the derivatives on the brane.
\begin{align}
    \partial_x^{2n}J|_{\partial \mathcal{M}}=\partial_x^{2n+1}s_1|_{\partial \mathcal{M}}=\partial_x^{2n+1}B|_Q=0\,,\quad n\in N\,.
\end{align}
These ensure that the boundary condition on the brane 
\begin{align}
    J|_{\partial \mathcal{M}}=0,\quad \partial_x s_1|_{\partial \mathcal{M}}=0,\quad \partial_xB|_Q=0,\quad \partial_x A|_Q=0\,,\label{eq:branecondition}
\end{align}
are satisfied during time evolution.

From \eqref{eqS} to \eqref{eqA}, we further obtain 
\begin{align}
    \partial_x^{2n+1}\Sigma|_Q= \partial_x^{2n}F|_Q=\partial_x^{2n+1}d_+\Sigma|_Q=\partial_x^{2n+1}d_+B|_Q=\partial_x^{2n+1}A|_Q=0\,,
\end{align}
which implies 
\begin{align}
    \partial_x^{2n+1}a_3|_{\partial \mathcal{M}}=\partial_x^{2n+1}b_3|_{\partial \mathcal{M}}=
    \partial_x^{2n+1}\dot{B}|_Q=
    \partial_x^{2n+1}\dot{S}|_Q=0\, .
\end{align}
Consequently, 
\begin{align}
    \partial_x^{2n}\dot{J}|_{\partial \mathcal{M}}=\partial_x^{2n+1}\dot{s}_1=0|_{\partial \mathcal{M}}\,,
\end{align}
proving that the boundary conditions  \eqref{eq:sumbranecond} and \eqref{eq:addcons} hold throughout the evolution. This corresponds to a mirror-reflecting boundary condition on the brane. 

\subsection{Time Evolution\label{time evolution}}
By solving the equations of motion, we  determine the time evolution of the geometry and study the dynamical behavior of the boundary system. Below we outline how to solve those equations. The dynamical variables are chosen to be $J$, $s_1$ and $B$. For initial conditions, we set $J(x)=0$ and $B(x,u)=0$, then solve $s_1(x)$ such that the energy density matches the the desired profile given in Eq.~\eqref{E_ini} in the main text. With these quantities specified, we proceed with the following steps to solve the equations of motion.

\begin{enumerate}
    \item Given $B$ and $s_1$, we use~\eqref{eqS} to solve $\Sigma$.
    \item Given $B$, $s_1$, $J$ and $\Sigma$, we use~\eqref{eqF} to solve $F$. 
    \item Given $B$, $s_1$, $\Sigma$ and $F$, we can use \eqref{eqdpS} to solve $d_+S$. The boundary condition at the apparent horizon requires vanishing expansion,
    \begin{align}
        d_+S\big|_{u=u_h}=\frac{e^{-B}\left( F \Sigma\partial_xB  -\Sigma\partial_xF  +F^2 u^2 \Sigma '\right)}
        {2 \Sigma ^2}\,.
    \end{align}
    By fixing this horizon condition, the energy density can be extracted  from the asymptotic solution of $d_+S$. Note that in this evolution scheme, the energy density is not an independent dynamical variable.  
    \item Given $B$, $s_1$, $\Sigma$, $F$ and $d_+S$, we determine $d_+B$ using~\eqref{eqdpB}. 
    \item With all previous quantities known, \emph{i.e.}  $B$, $s_1$, $\Sigma$, $F$, $d_+S$ and $d_+B$, we solve for $A$ using~\eqref{eqA}. The boundary condition at the apparent horizon is obtained by solving \eqref{eqAh} with the boundary condition \eqref{eq:branecondition}.
    \item We then get $\dot{B}$ using the definition of $d_+$ and $d_+B$, and get $\dot{s}_1$ from the asymptotically solution of $A$ , and get $\dot{J}$ from energy-momentum tensor conservation.
    \item By employing fourth-order Runge-Kutta method for the first three time steps and then the fourth-order  Adams-Bashforth method for subsequent steps, we  compute $B(u,t+\delta t,x)$, $J(u,t+\delta t,x)$ and $s_1(u,t+\delta t,x)$, after which we repeat the entire procedure from step 1.
\end{enumerate}

We initialize our system as a locally boosted black brane.  At $t=0$, we must specify the profile of $J$, $s_1$ and $B$. First, we choose 
\begin{align}
    B=0,\quad J=0\,.
\end{align}
And the initial energy density is given in Eq. \eqref{E_ini} in the main text, from which we determine the initial profile of $s_1$ by solving \eqref{eqS}, \eqref{eqF}, and \eqref{eqdpS}. This completes the specification of initial conditions for solving the bulk equations.

Notably, our system possesses  the following scaling  symmetry: 
\begin{equation}
(u,t,x,y)\rightarrow\lambda(u,t,x,y)\,,\quad A\rightarrow\frac{1}{\lambda^2}A\,,\quad\Sigma\rightarrow\frac{1}{\lambda^2}\Sigma\,,
\end{equation}
which induces corresponding transformations of the  boundary quantities 
\begin{equation}\label{scaling}
 L\rightarrow{\lambda}L\,,\quad {E}\rightarrow\frac{1}{\lambda^3}{E}\,.
\end{equation}
Due to this scaling symmetry, we characterize our system using the dimensionless ratios
\begin{align}
    \Delta=\frac{E_R-E_L}{E_L}\,,\quad\Bar{L}=L\left(\frac{E_R+E_L}{2}\right)^{1/3}\,,
\end{align}
where $\Delta$ measures the energy imbalance and $\Bar{L}$
parameterize the scaled system size.
\\

\subsection{Entropy production and energy conservation}\label{energy conservation}

In an isolated system, entropy cannot decrease while the total energy remains  conserved. As demonstrated in Fig. \ref{fig:NESS_energy_entropy_t} in the main text, and further illustrated in Fig. S2 
below, we observe entropy growth in all systems under consideration.
\begin{figure}[h]
    \centering
    \includegraphics[width=0.32\textwidth]{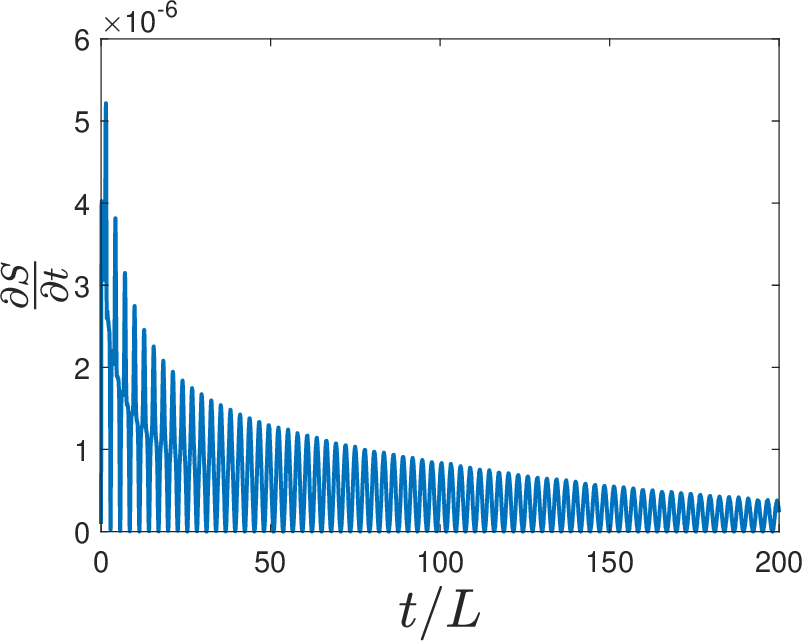}
    \includegraphics[width=0.32\textwidth]{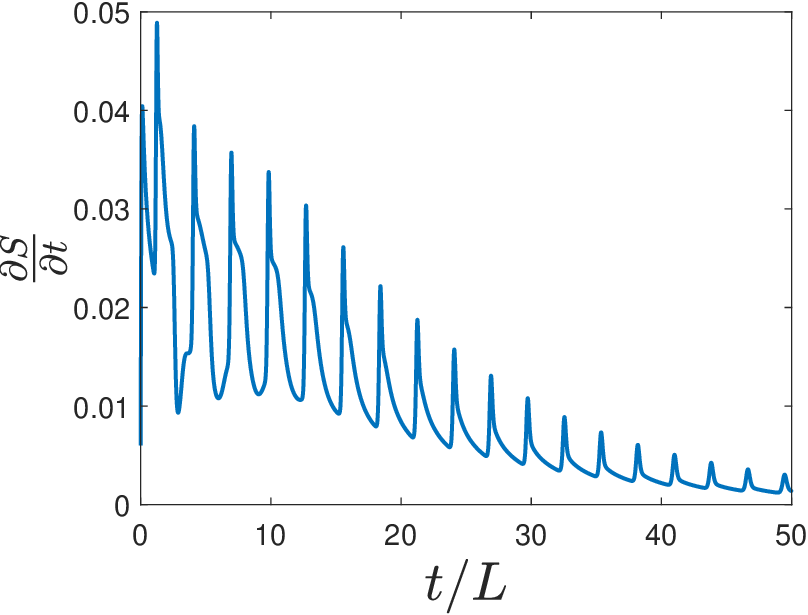}
    \includegraphics[width=0.32\textwidth]{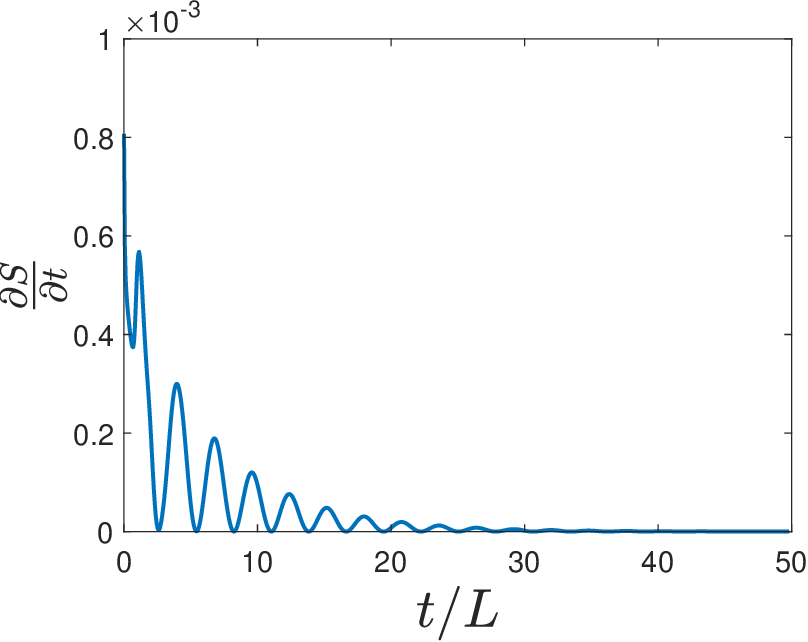}
    \caption{The time evolution of the entropy (area of apparent horizon) change rate. \textbf{\em Left panel:} $\Delta=0.02$, $\bar{L}=100$, NESS recurrence. \textbf{\em Middle panel:} $\Delta=3$, $\bar{L}=50$, confined shock wave. \textbf{\em Right panel:} $\Delta=0.08$, $\bar{L}=5$, oscillatory decay.}
    \label{fig:entropy_two}
\end{figure}

The total energy of the system remains conserved throughout the evolution, as clearly  shown in Fig.  S3. This can also be used to check the accuracy of our numerical computation.
\begin{figure}[h]
    \centering
    \includegraphics[width=0.32\textwidth]{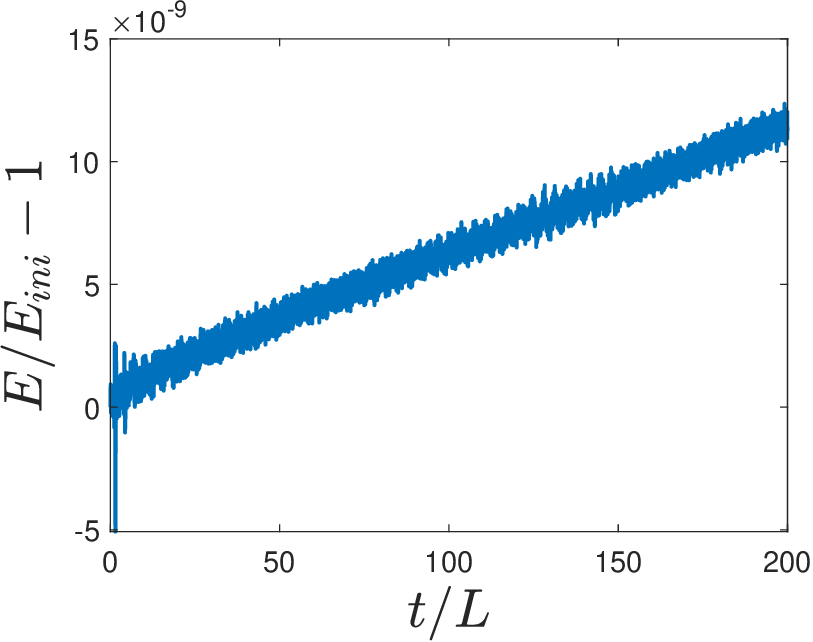}
    \includegraphics[width=0.32\textwidth]{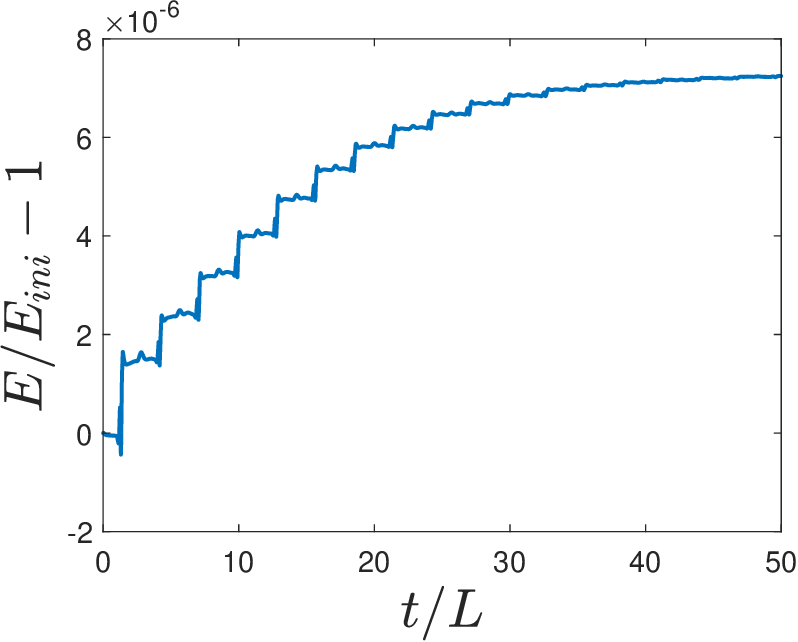}
    \includegraphics[width=0.32\textwidth]{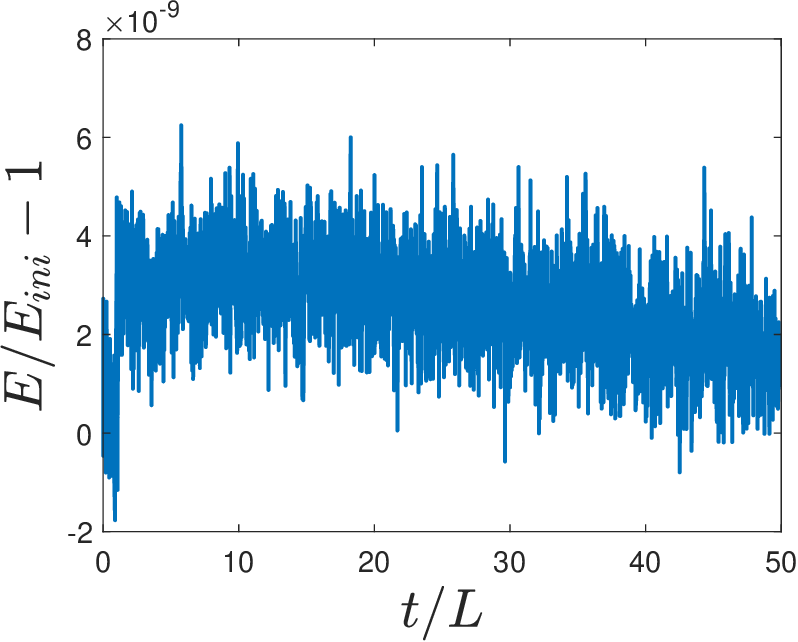}
    \caption{Demonstration for the energy conservation during the time evolution. $E_{ini}$ is initial total energy. \textbf{\em Left panel:} $\Delta=0.02$, $\bar{L}=100$, recurrent NESS . \textbf{\em Middle panel:} $\Delta=3$, $\bar{L}=50$, confined shock wave. \textbf{\em Right panel:} $\Delta=0.08$, $\bar{L}=5$, oscillatory decay. }
    \label{fig:conservE}
\end{figure}

\subsection{Perturbations analysis}\label{pertanalusis}

In the final stage of thermalization, all physical quantities in the system approach global equilibrium via oscillation decay, where the oscillation  frequency is governed by the QNMs of the bulk black brane. To see this, we preform a  perturbation analysis of the system. The perturbation of metric is given by 
\begin{align}
    \delta g_{\mu\nu}=\begin{bmatrix}
\delta g_{tt}&0 & \delta g_{tx} & 0  \\
0 & 0 &0 &0\\
\delta g_{tx} & 0 &\delta g_{xx} &0\\
0 & 0  &0 &\delta g_{yy}
\end{bmatrix}\,.
\end{align}
We then decompose the perturbation into Fourier modes 
\begin{align}
    \delta g_{\mu\nu}=\sum_{n=1}^{\infty}\delta \hat{g}_{\mu\nu}(u)e^{-i\omega t+ik_nx}\,,
\end{align}
where $k_n=\frac{n\pi}{2L}$ is discrete wave-number due to the finite system size. From this, we obtain  the eigenfrequencies of the system, as shown in Fig. S4.

\begin{figure}[h]
    \centering
    \includegraphics[width=0.43\textwidth]{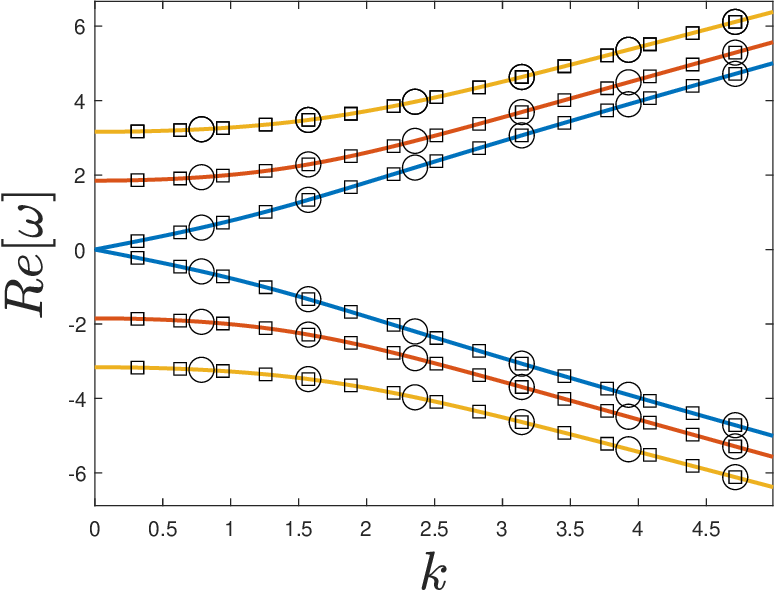}\quad\quad
    \includegraphics[width=0.43\textwidth]{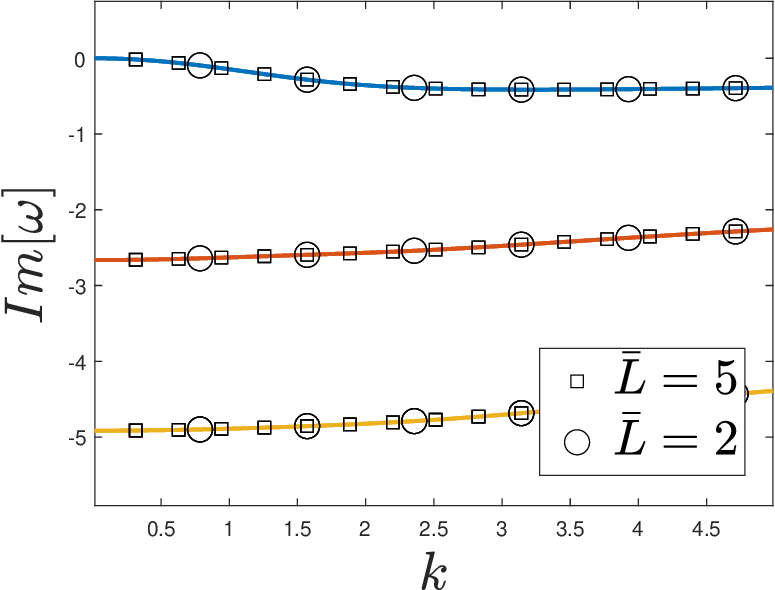}
    \caption{The first three QNMs versus wave-number for different size. Square and circle are the modes allowed for different size $\bar{L}$.}
    \label{fig:pert}
\end{figure}

With the QNMs, we can analyze the behavior during the final stage of thermalization. As an example, we consider the results shown in Fig. \ref{fig:qnm_t}. 
The early time behavior can be well fitted for $k_1$ and $k_3$, as illustrated in Fig. S5. In contrast, even modes such as $k_2$
  exhibit an initial increase due to nonlinear effects.  However, as the amplitudes of different $k_n$ decay, higher modes become significantly smaller than the first  QNM. This leads to a crossover where higher modes enter a forced oscillation stage, with the first QNM acting as the driving source. In this stage, the driven oscillation frequency of the higher modes becomes $n\,\omega_1$. 
 
\begin{figure}[h]
    \centering
    \includegraphics[width=0.45\textwidth]{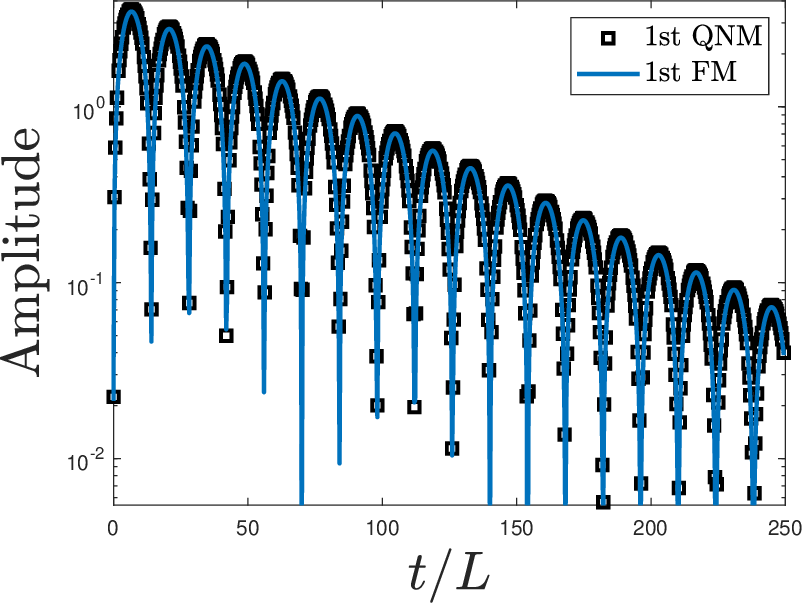}\quad\quad
    \includegraphics[width=0.45\textwidth]{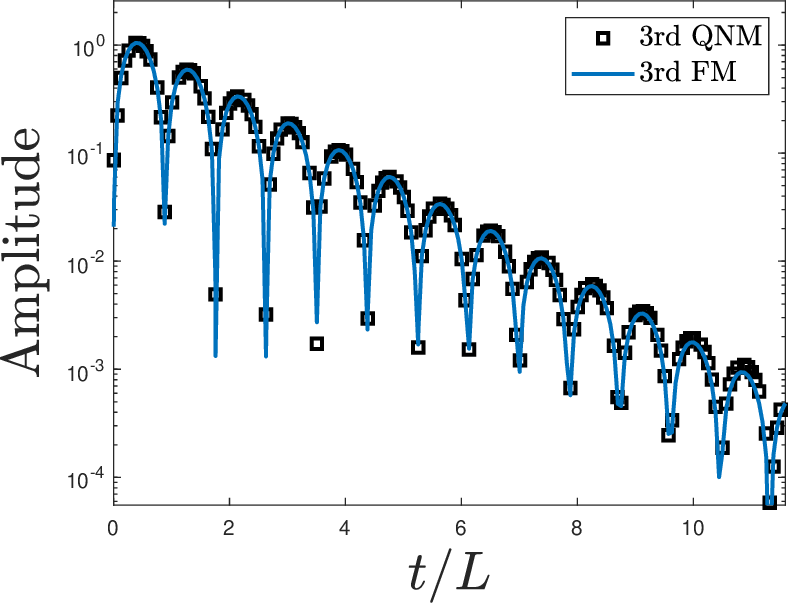}
    \caption{Comparison between the linear QNMs and the FMs extracted directly from the time evolution. We find a quantitative agreement. Here $\Delta=0.08, \bar{L}=5$.}
    \label{fig:oddn}
\end{figure}

Whether a mode exhibits forced oscillation or continues to follow its intrinsic quasi-normal behavior depends on the imaginary parts of the frequencies: if  $\text{Im}[n\,\omega_1]>\text{Im}[\omega_n]$, the  $n$-th mode will eventually become forced; if $\text{Im}[n\,\omega_1]<\text{Im}[\omega_n]$, it will continue to decay according to its own QNM until  the end. 

Within the first-order hydrodynamics approximation, $\omega_n=n \text{Re}[\omega_1]+in^2 \text{Im}[\omega_1]$. Therefore 
$\text{Im}[n\omega_1]=n \text{Im}[\omega_1]>\text{Im}[\omega_n]=n^2\text{Im}[\omega_1]$, and the system shows a crossover from
intrinsic quasi-normal decay to driven oscillation. However, for larger $n$ or small $\bar{L}$, the imaginary part of quasi-normal frequencies exhibits a  plateau (as seen in Fig. S4), causing higher QNMs to dominate again. 

For $\Delta=0.08, \bar{L}=5$, we find  $\text{Im}[n\omega_1]>\text{Im}[\omega_n]$ for $n=2,3$ modes. Consequently, as shown in Fig. S6, the driven frequency prediction agrees well with the full numerical time evolution.

\begin{figure}[h]
    \centering
    \includegraphics[width=0.45\textwidth]{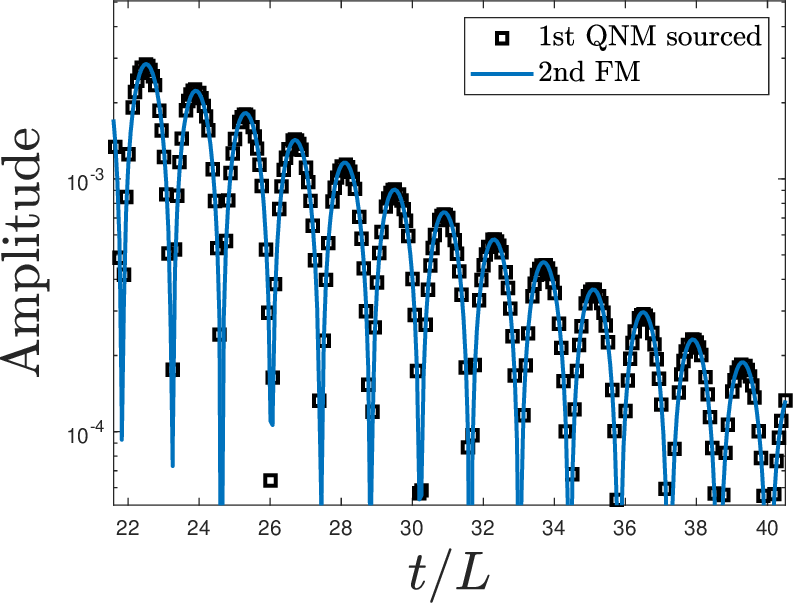}\quad\quad
    \includegraphics[width=0.45\textwidth]{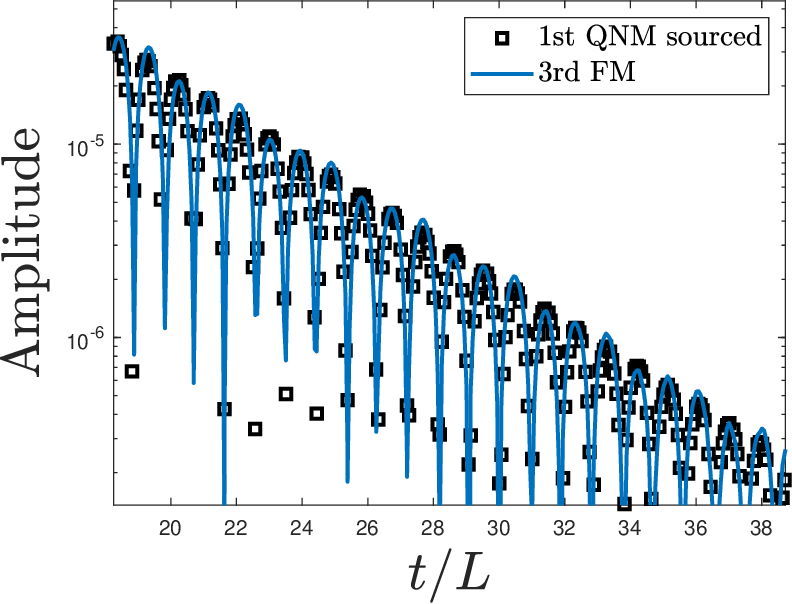}
    \caption{In the forced oscillation stage, the frequency of higher mode can be described by $n\omega_1$. We have chosen $\Delta=0.08, \bar{L}=5$.}
    \label{fig:qnm1drive}
\end{figure}

\subsection{Error analysis}\label{erranalysis}

During time evolution, we do not directly evolve $\Sigma$. Instead, we use $d_+\Sigma$ in the time evolution. 
To verify the numerical evolution, we can examine 
the time derivative of $\Sigma$, which can be computed in two distinct ways:  
\begin{align}
    \dot{\Sigma}_1(t)&=\frac{\Sigma(t+\delta t)-\Sigma(t-\delta t)}{2\delta t}\,,\\
    \dot{\Sigma}_2(t)&=d_+\Sigma(t)+\frac{u^2}{2}A(t)\Sigma(t)\,.
\end{align}
Therefore, the first quantity we use to estimate the numerical error is
\begin{align}
    err_1=\frac{|\dot{\Sigma}_1(t)-\dot{\Sigma}_2(t)|}{max(|\dot{\Sigma}_1|)}\,.
\end{align}
The errors in different parameters are presented in Fig. S7.
\begin{figure}[h]
    \centering
    \includegraphics[width=0.32\textwidth]{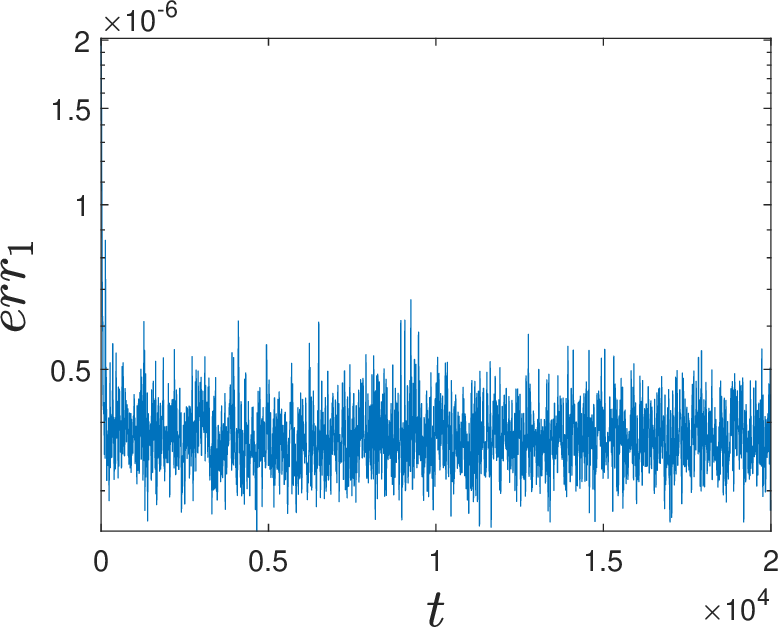}
    \includegraphics[width=0.32\textwidth]{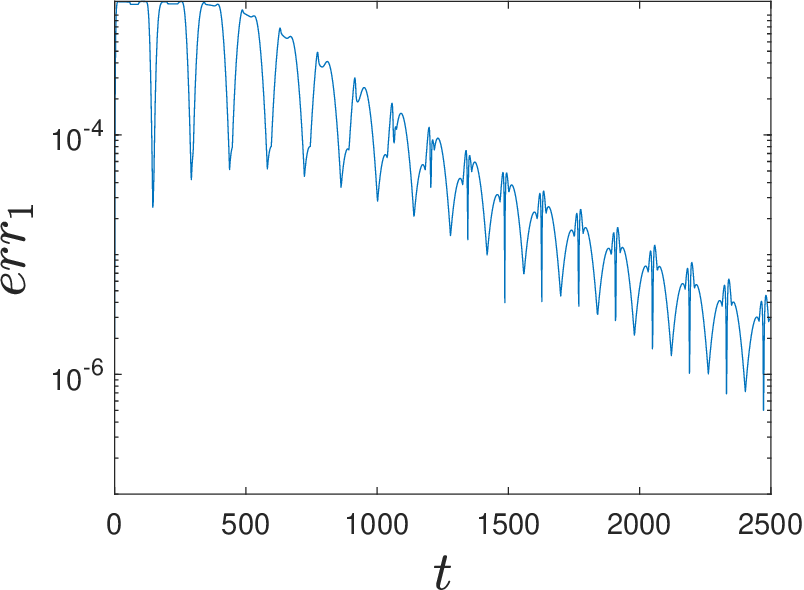}
    \includegraphics[width=0.32\textwidth]{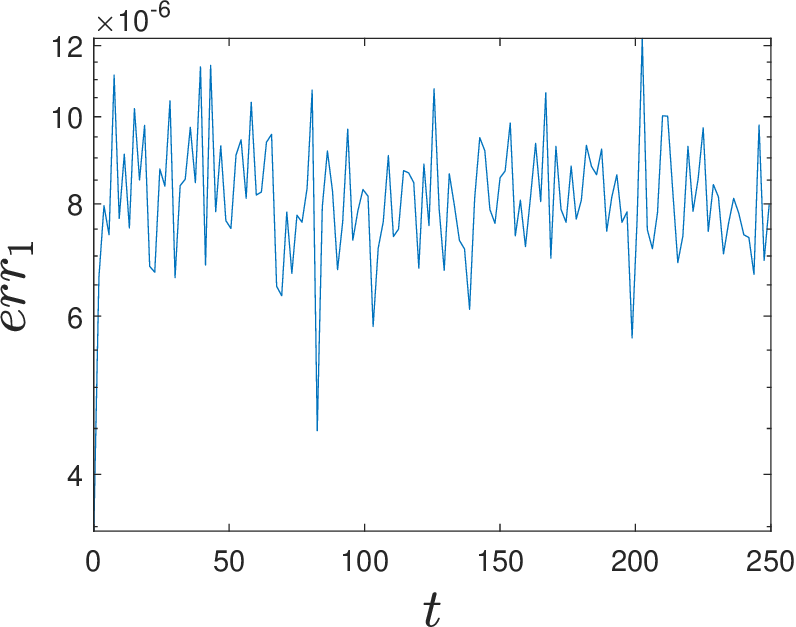}
    \caption{Relative error in $\dot{\Sigma}$. \textbf{\em Left panel:} $\Delta=0.02$, $\bar{L}=100$.  \textbf{\em Middle panel:} $\Delta=3,\ \bar{L}=50$ . \textbf{\em Right panel:} $\Delta=0.08,\ \bar{L}=5$. } 
    \label{fig:err11}
\end{figure}

An alternative approach involves other equations of motion that are not directly solved in our evolution scheme. For example, one equation containing a second-order time derivative is 
\begin{align}
&-2 F^2 \Sigma ^2 u^4 A''+2 \Sigma ^2 u^2 A' \left(F^2 u^2 B'+2 e^B \dot{\Sigma } \Sigma -B_x F+F_x-2 F^2 u\right)
+2 A F^2 \Sigma ^2 u^4 B''\notag\\
&+2 \Sigma  u^2 B' \left(A \left(e^B (-\dot{B}) \Sigma ^3+B_x F \Sigma -2 \Sigma _x F+2 F^2 \Sigma  u\right)+2 A F^2 u^2 \Sigma '+F \Sigma  (\dot{F}-A_x)\right)-2 A F^2 \Sigma ^2 u^4 \left(B'\right)^2\notag\\
&-8 A e^B \Sigma ^3 u^2 \dot{\Sigma }'-2 A F \Sigma ^2 u^2 B_x'+2 A B_x \Sigma ^2 u^2 F'-2 A \Sigma ^2 u^2 F_x'+4 A F \Sigma  u^2 \Sigma _x'-4 A \Sigma _x F u^2 \Sigma '\notag\\
&-4 A F^2 \Sigma  u^4 \Sigma ''+4 A F^2 u^4 \left(\Sigma '\right)^2-8 A F^2 \Sigma  u^3 \Sigma '+4 e^B \ddot{\Sigma } \Sigma ^3-4 e^B \dot{A} \Sigma ^3 u^2 \Sigma '+e^B \dot{B}^2 \Sigma ^4-2 A_{xx} \Sigma ^2\notag\\
&+2 \dot{F}_x \Sigma ^2-2 F \Sigma ^2 u^2 \dot{F}'-2 \dot{F} B_x \Sigma ^2+4 F \Sigma ^2 u^2 A_x'+2 A_x B_x \Sigma ^2-2 A_x \Sigma ^2 u^2 F'=0\,.\label{eq:cons2order}
\end{align}
We use the evolution data and finite difference method to compute the time derivative of $\Sigma$, $F$ and $A$ and substitute into \eqref{eq:cons2order}. To check the degree to which the equation is satisfied, we compute the absolute value of each term on the left of  \eqref{eq:cons2order} and add up. Denoting the maximum value as $M_{eq}$, the error is defined as 
\begin{align}
  err_2=\eqref{eq:cons2order}/M_{eq}.  
\end{align}
The relative errors of the equation for different cases are shown in Fig. S8. 
\begin{figure}[ht]
    \centering
    \includegraphics[width=0.32\textwidth]{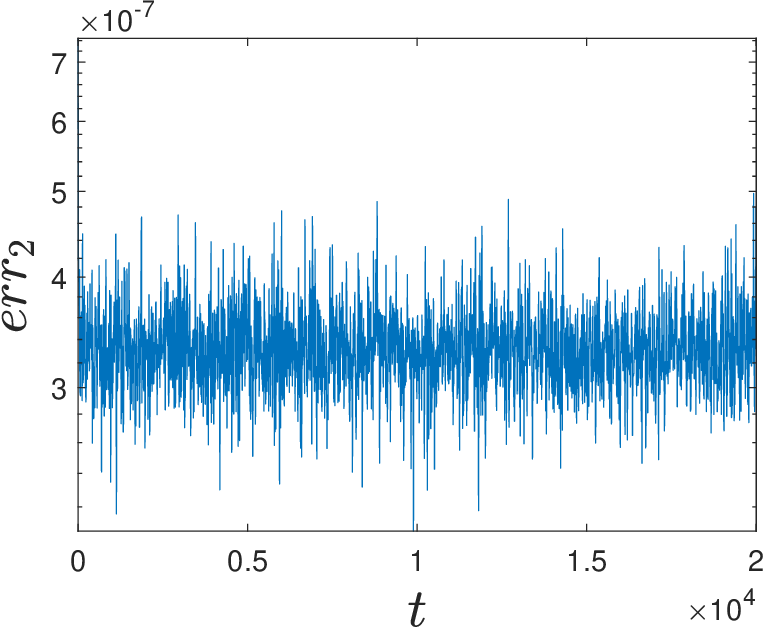}
    \includegraphics[width=0.32\textwidth]{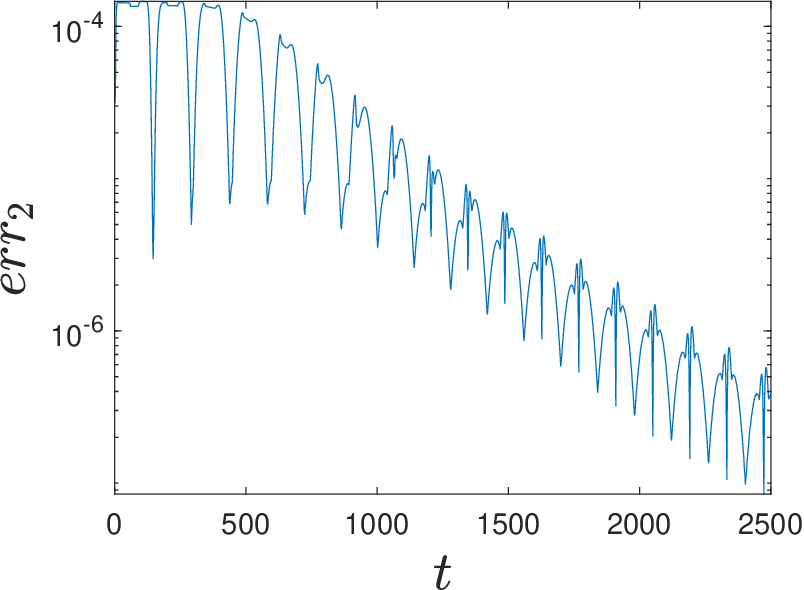}
    \includegraphics[width=0.32\textwidth]{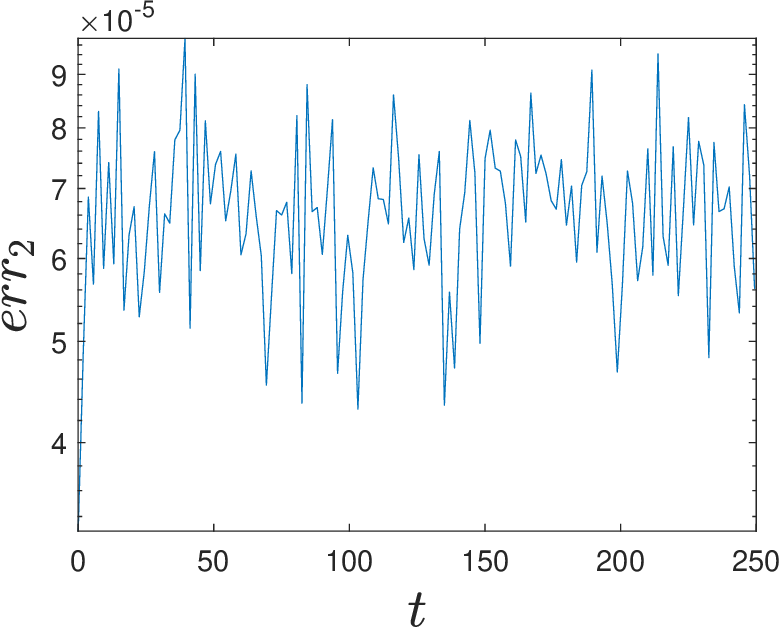}
    \caption{Relative error in Eq. \eqref{eq:cons2order}. \textbf{\em Left panel:} $\Delta=0.02$, $\bar{L}=100$.  \textbf{\em Middle panel:} $\Delta=3,\ \bar{L}=50$ . \textbf{\em Right panel:} $\Delta=0.08,\ \bar{L}=5$.  }
    \label{fig:err2}
\end{figure}

%

%

%
%


\end{document}